\documentclass[twocolumn,iop,revtex4]{openjournal}
\usepackage{natbib} 
\usepackage[dvipsnames]{xcolor}
\usepackage{aas_macros} 
\usepackage{amssymb}
\usepackage{amsmath}
\usepackage[title]{appendix}
\usepackage{hyperref}	
\hypersetup{colorlinks=true,linkcolor=blue,citecolor=blue,filecolor=blue,urlcolor=blue}
\usepackage[caption=false]{subfig}
\usepackage{ulem} 
\usepackage{xifthen}

\newcommand{\enzo}{\texttt{Enzo~}}
\newcommand{\enzoc}{\texttt{Enzo}}
\newcommand{\yt}{\texttt{yt~}}

\newcommand{\cloudy}{\texttt{Cloudy~}}
\newcommand{\grackle}{\texttt{Grackle-2.1~}}

\newcommand{\kms} {km $\rm{s^{-1}}$}

\newcommand{\msolar} {$\rm{M_{\odot}}~$}
\newcommand{\msolarc} {$\rm{M_{\odot}}$}
\newcommand{\msolaryr} {$\rm{M_{\odot}~yr^{-1}}~$}
\newcommand{\msolaryrc} {$\rm{M_{\odot}~yr^{-1}}$}

\newcommand{\zsolarc} {$\rm{Z_{\odot}}$}

\newcommand{\molH} {$\rm{H_2}$~}

\newcommand{\JLW} {J$_{\rm LW}$}

\newcommand{\rarepeak} {\textit{Rarepeak~}}
\newcommand{\rarepeakc} {\textit{Rarepeak}}
\newcommand{\normal} {\textit{Normal~}}

\newcommand{\void} {\textit{Void~}}

\newcommand{\ha} {\texttt{HaloA~}}
\newcommand{\hb} {\texttt{HaloB~}}
\newcommand{\hac} {\texttt{HaloA}}
\newcommand{\hbc} {\texttt{HaloB}}

\newcommand{\change}[2][]{%
\ifthenelse{\isempty{#2}}{{\color{ForestGreen}{#1}}}%
{{\color{RedOrange}\sout{#1}}{\color{ForestGreen}{ #2}}}%
}

\begin{document}
\title[Very Massive Star Formation in Early Galaxies]{The Formation of Very Massive Stars in Early Galaxies and Implications for Intermediate Mass Black Holes}
\author{John A. Regan$^{1,2,*}$}
\thanks{$^*$E-mail:john.regan@mu.ie, Royal Society - SFI University Research Fellow}
\author{John H. Wise$^{3}$}
\author{Tyrone E. Woods$^{4}$}
\author{Turlough P. Downes$^{2}$}
\author{Brian W. O'Shea$^{5,6,7, 8}$}
\author{Michael L. Norman$^9$}

\affiliation{$^1$Department of Theoretical Physics, Maynooth University, Maynooth, Ireland}
\affiliation{$^2$Centre for Astrophysics \& Relativity, School of Mathematical Sciences, Dublin City University, Glasnevin, D09 W6Y4, Ireland}
\affiliation{$^3$Center for Relativistic Astrophysics, Georgia Institute of Technology, 837 State Street, Atlanta, GA 30332, USA}
\affiliation{$^4$National Research Council of Canada, Herzberg Astronomy \& Astrophysics Research Centre, 5071 West Saanich Road, Victoria, BC V9E 2E7, Canada}
\affiliation{$^5$Department of Computational Mathematics, Science, and Engineering, Michigan State University, MI, 48823, USA}
\affiliation{$^6$Department of Physics and Astronomy, Michigan State University,MI, 48823, USA}
\affiliation{$^7$Joint Institute for Nuclear Astrophysics - Center for the Evolution of the Elements, USA}
\affiliation{$^8$National Superconducting Cyclotron Laboratory, Michigan State, University, MI, 48823, USA}
\affiliation{$^9$Center for Astrophysics and Space Sciences, University of California, San Diego, 9500 Gilman Dr, La Jolla, CA 92093}


\begin{abstract}
  We investigate the ab-initio formation of super-massive stars in a pristine atomic cooling halo.
  The halo is extracted from a larger self-consistent parent simulation. The halo remains metal-free
  and star formation is suppressed due to a combination of dynamical heating from mergers and a
  mild ($J_{\rm LW} \sim 2 - 10 \ J_{21}$(z)) Lyman-Werner (LW) background.
  We find that more than 20 very massive stars form with stellar
  masses greater than 1000 \msolarc. The most massive star has a stellar mass of over 6000  \msolarc.
  However, accretion onto all stars declines significantly after the first $\sim$ 100 kyr of evolution
  as the surrounding material is accreted and the turbulent nature of the gas causes the stars to
  move to lower density regions. We post-process the impact of ionising radiation from the stars and
  find that ionising radiation is not a limiting factor when considering SMS formation
  and growth. Rather the birth environments are highly turbulent and a steady accretion flow is not
  maintained within the timescale (2 Myr) of our simulations. As the massive stars end their lives
  as direct collapse black holes this will seed these embryonic haloes with a population of black
  holes with masses between approximately 300 \msolar and 10,000 \msolarc.
  Afterwards they may sink to the centre of the haloes, eventually coalescing to form larger
  intermediate mass black holes whose in-situ mergers will be detectable by LISA. 
\end{abstract}


\section{Introduction} \label{Sec:Introduction}
Supermassive stars (SMSs) with masses between $10^4$ and $10^5$ \msolar have, over the past few decades, been invoked \citep{Rees_1978, Begelman_1978, Begelman_2006,
  Begelman_2008, Latif_2016a, Woods_2018} as an intermediate phase to explain the existence of supermassive black holes (SMBHs) at the centres of massive galaxies \citep{Fan_06, Kormendy_2013}.
The pathways to forming a SMBH are thus far unknown with a number of theoretical models proposing to explain their existence. \\
\indent Perhaps the
simplest explanation is to invoke the black holes left over from the first 
generation of stars as seeds for SMBHs. The first generation of (metal-free) 
stars are referred to as Population III (PopIII) stars and according to 
current theoretical models \citep[e.g.][]{Turk_2009, Clark_2008, Hirano_2014, Stacy_2016} the initial mass function should be
top heavy with a characteristic mass of tens of solar masses. However, PopIII remnant black holes are expected to form in low
density environments \citep{Whalen_2004, OShea_2005b, Milosavljevic_2009} and are not expected to accrete substantially, at least not 
initially \citep{Alvarez_2009, Smith_2018}. PopIII stars are therefore not seen as good candidates to explain the existence
of SMBHs without invoking super-Eddington accretion scenarios which can boost their initial 
seed masses by an order of magnitude or more over a short period
\citep{Lupi_2014, Pacucci_2015a, Sakurai_2016a,Inayoshi_2016, Pacucci_2017, Inayoshi_2018}.\\
\indent SMSs provide an alternative path to forming a SMBH by giving the seed black hole a
head-start compared to a black hole formed from a PopIII remnant. Under a SMS formation scenario
the accretion rate onto the protostar must exceed a critical threshold thought to be around
0.001 \msolaryr \citep{Haemmerle_2017}. When
this threshold accretion rate is reached and maintained the stellar radius inflates reducing
its surface temperature to approximately 5000 K
and making the star resemble a red giant star
\citep{Omukai_2003, Hosokawa_2012, Hosokawa_2013, Woods_2017}. However, the SMS must continue to
accrete above this threshold rate. If the accretion rate falls below the critical
rate, for a time exceeding the Kelvin-Helmholtz time \citep{Sakurai_2016}, the star contracts to the main sequence and becomes a hyper-luminous PopIII
star with a mass set approximately by the mass at which the accretion rate dropped.
When the accretion rate is maintained the star grows rapidly but emits
only weak radiative feedback with the spectrum of the emitted radiation peaking below
the hydrogen ionisation limit \citep{Woods_2018}. \\
\indent As discussed, the key requirement for
forming a SMS is that the mass accretion rate onto the star exceeds approximately 0.001 \msolaryrc,
however, a sufficient baryon reservoir is also required and furthermore the metallicity of the gas
being accreted should be below $10^{-3}$ \zsolarc.  Detailed high resolution simulations have found that gas that has been enriched above this threshold fragments into lower mass stars which do not converge to a single object and in this case the formation of a SMS is suppressed \citep[e.g.][]{Chon_2020}. For these reasons
metal-poor (i.e., Z $\lesssim 10^{-3}$ \zsolarc) atomic cooling haloes are seen as the most 
promising candidates in which to form SMSs. Haloes which have higher levels of 
metal-enrichment (Z $> 10^{-3}$ \zsolarc) may also be viable candidates for 
SMS formation in the early universe if metal mixing is inhomogeneous \citep{Regan_2020a}. Atomic cooling
haloes which provide the above requirements for SMS formation were recently investigated by \cite{Wise_2019}
and \cite{Regan_2020}. In particular \cite{Wise_2019} found that the combination 
of a mild Lyman-Werner background combined with the impact of dynamical heating effects due to
minor and major mergers can suppress star formation until a halo crosses the atomic cooling threshold.
These haloes are therefore predominantly metal-poor (with any metal enrichment coming externally), have large 
baryon reservoirs and suppressed \molH content due to the LW radiation fields. 
In this study we build on the previous works cited above. \\
\indent The goal of this study is to model the formation and evolution of (super-)massive star
formation in haloes that are exposed to moderate LW backgrounds, which when combined with
the effects of dynamical heating can suppresses star formation below the atomic cooling limit.
  To pursue this research we re-simulate
  two haloes from the original Renaissance simulations using the zoom technique.
  We designate
  these haloes as \ha and \hbc. Both haloes were chosen as they exhibited near isothermal
  collapse of their inner core in the original Renaissance datasets as shown by \cite{Regan_2020}.
  They were therefore identified as among the most promising candidates for SMS formation. Both
  haloes were exposed to moderate levels of LW radiation from nearby radiation sources as well as
  constant mergers which dynamically heated the gas within the haloes. To further understand the
  impact of the LW field we re-simulate \hb with no LW radiation in this work. This is done to
  determine if a halo can remain star-free due to only dynamical heating effects or if the LW field 
  remains a critical component. \hac, on the other-hand, is re-simulated with a LW background
  composed of both local source contributions and background contributions. \\
  \indent In the zoom simulations, we find that \ha forms stars with masses greater than
  6000 \msolar but that the accretion rate onto individual proto-stars always declines as the star's
  immediate gas supply is depleted. In the re-simulation of \hbc, without a LW field, we find that
  the halo undergoes premature collapse (compared to the original case where a LW field of
  \JLW $\sim 2$ J$_{21}$\footnote{J$_{21}$\ is shorthand for $1 \times 10^{-21} \ \rm{ erg\ cm^{-2}\ s^{-1}\ Hz^{-1}\ sr^{-1}}$} existed). In \hb the most massive star in the halo has a mass of
  approximately 173 \msolarc. The impact of ionising radiation is not considered in these
  simulations but post-processing of the stellar feedback using \cloudy \citep{Ferland_2017} is
  instead used to gauge the likely impact of ionising sources, particularly for \ha which forms a
  number of hyper-luminous PopIII stars. \\
  \indent The paper is laid out as follows: In \S \ref{Sec:Methods} we very briefly review the
  original Renaissance simulations as well as discussing the zoom-in simulations. In \S
  \ref{Sec:Results} we analyse the results of the zoom-in simulations. In \S \ref{Sec:GW} we
  discuss the implications of the results and the connection with upcoming gravitational wave
  observatories. In \S \ref{Sec:Discussion} we summarize our results and outline
  our conclusions.

\section{Methods} \label{Sec:Methods}

\subsection{Renaissance Simulation Suite} \label{Sec:Renaissance}
\enzo has been extensively used to study the formation of structure in the early universe
\citep{Abel_2002, OShea_2005b, Turk_2012, Wise_2012b, Wise_2014, Regan_2015, Regan_2017}.
\enzo includes a ray tracing scheme to follow the propagation of radiation from
star formation and black hole formation \citep{WiseAbel_2011} as well as a detailed multi-species
chemistry model that tracks the formation and evolution of nine species \citep{Anninos_1997,
  Abel_1997}. In particular the photo-dissociation of \molH is followed, which is a critical
ingredient for determining the formation of the first metal-free stars \citep{Abel_2000}.\\
\indent The original Renaissance simulations \cite{Xu_2013, Xu_2014, OShea_2015} were carried out
on the Blue Waters supercomputer using the adaptive mesh refinement
code \enzo\citep{Enzo_2014, Enzo_2019}\footnote{https://enzo-project.org/}.
The datasets that formed the basis for this study were originally derived from a simulation of the
universe in a 40 Mpc on the side box using the WMAP7 best fit cosmology \citep{Komatsu_2011}.
For more details on the Renaissance simulation suite see \cite{Chen_2014}. Here we outline only
the details relevant to this study for brevity. The simulation suite was broken down into
three separate regions, namely the \rarepeakc, \normal and \void regions. Each region was simulated
with an effective initial resolution of $4096^3$ grid cells and particles giving a maximum dark matter
particle mass resolution of $2.9 \times 10^4$ \msolarc. Further refinement was allowed throughout
each region up to a maximum refinement level of 12, which corresponded to 19 pc comoving spatial
resolution. Given that the regions focus on different
 overdensities each region was evolved forward in time to different epochs. The \rarepeak region,
 being the most overdense and hence the most computationally demanding at earlier times, was run
 until $z = 15$. The \normal region ran until $z = 11.6$, and the \void region ran until $z = 8$.
 In all of the regions the halo mass function was very well resolved down to M$_{\rm halo} \sim 2
 \times 10^6$ \msolarc. \\
 \indent  As noted already in \S \ref{Sec:Introduction}, in \cite{Wise_2019} we examined two
 metal-free and star-free haloes from the final output of the \rarepeak simulation and re-simulated
 those two haloes at significantly higher resolution (maximum spatial resolution,
 $\Delta x \sim 60$ au) until the point of collapse. This re-simulation allowed us to investigate
 the evolution of the inner halo and the mass distribution of the clumps formed. However, no star
 formation prescription was employed during this re-simulation. In \cite{Regan_2020} we subsequently
 investigated the occurrence of metal-free and star-free atomic cooling haloes across all of the
 Renaissance datasets. We found a total of 79 such haloes in the
 \rarepeak outputs and three such haloes in the \normal outputs. None were found in the \void outputs.
 Of the 79 haloes which were metal-free and star free above the atomic cooling limit four haloes showed almost ideal isothermal collapse (see Figure 6. in \cite{Regan_2020}) consistent with what has previously been identified
 as ideal conditions for forming SMSs \citep{Inayoshi_2014, Becerra_2015, Latif_2016a,
   Regan_2017, Chon_2017b, Regan_2018b}. Of those haloes which collapsed isothermally we then
 selected two haloes for re-simulation in this study. \\

\begin{figure*}
\centering
\begin{minipage}{175mm}      \begin{center} 
\centerline{
\includegraphics[width=0.52\textwidth]{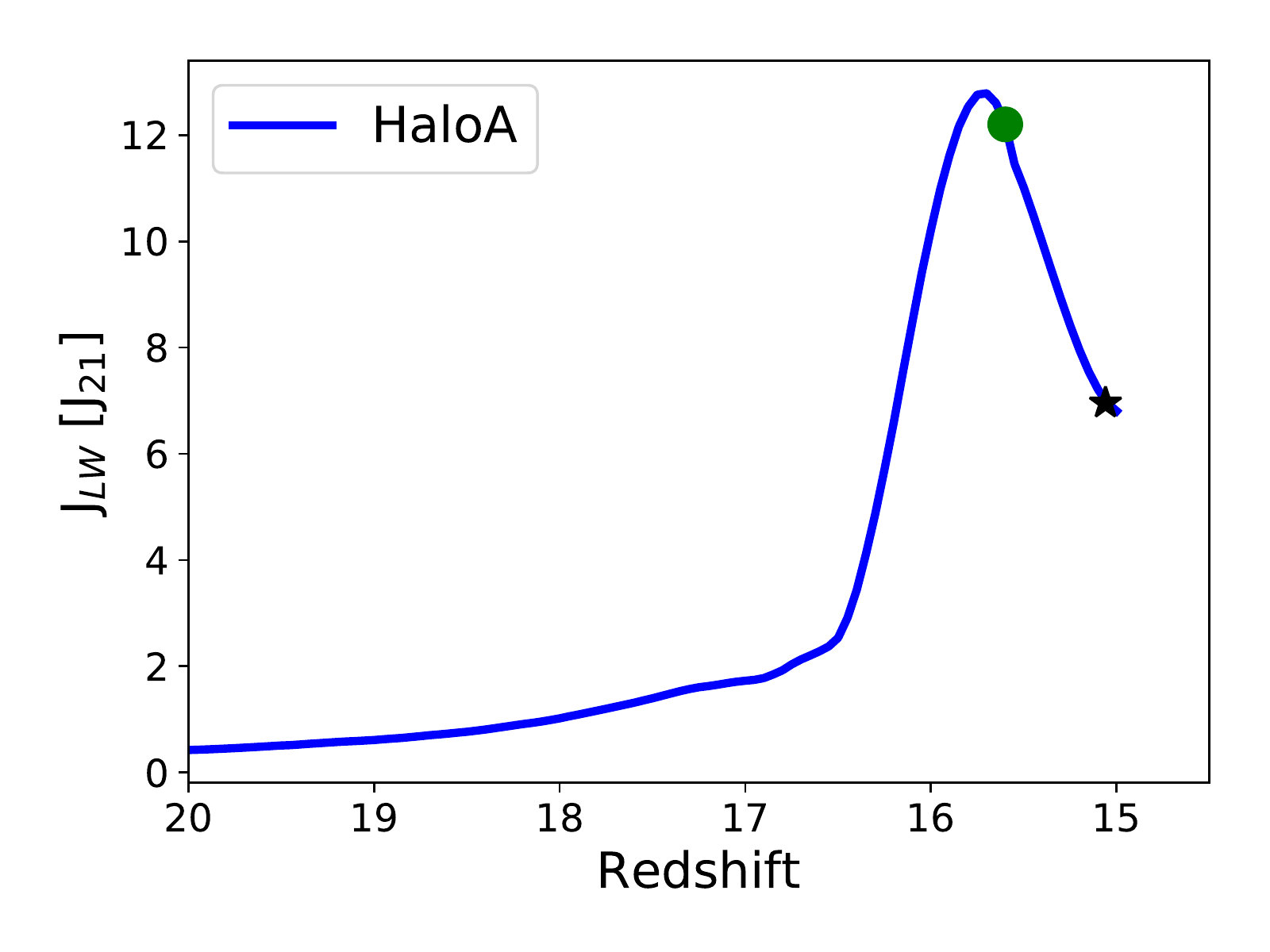}
\includegraphics[width=0.52\textwidth]{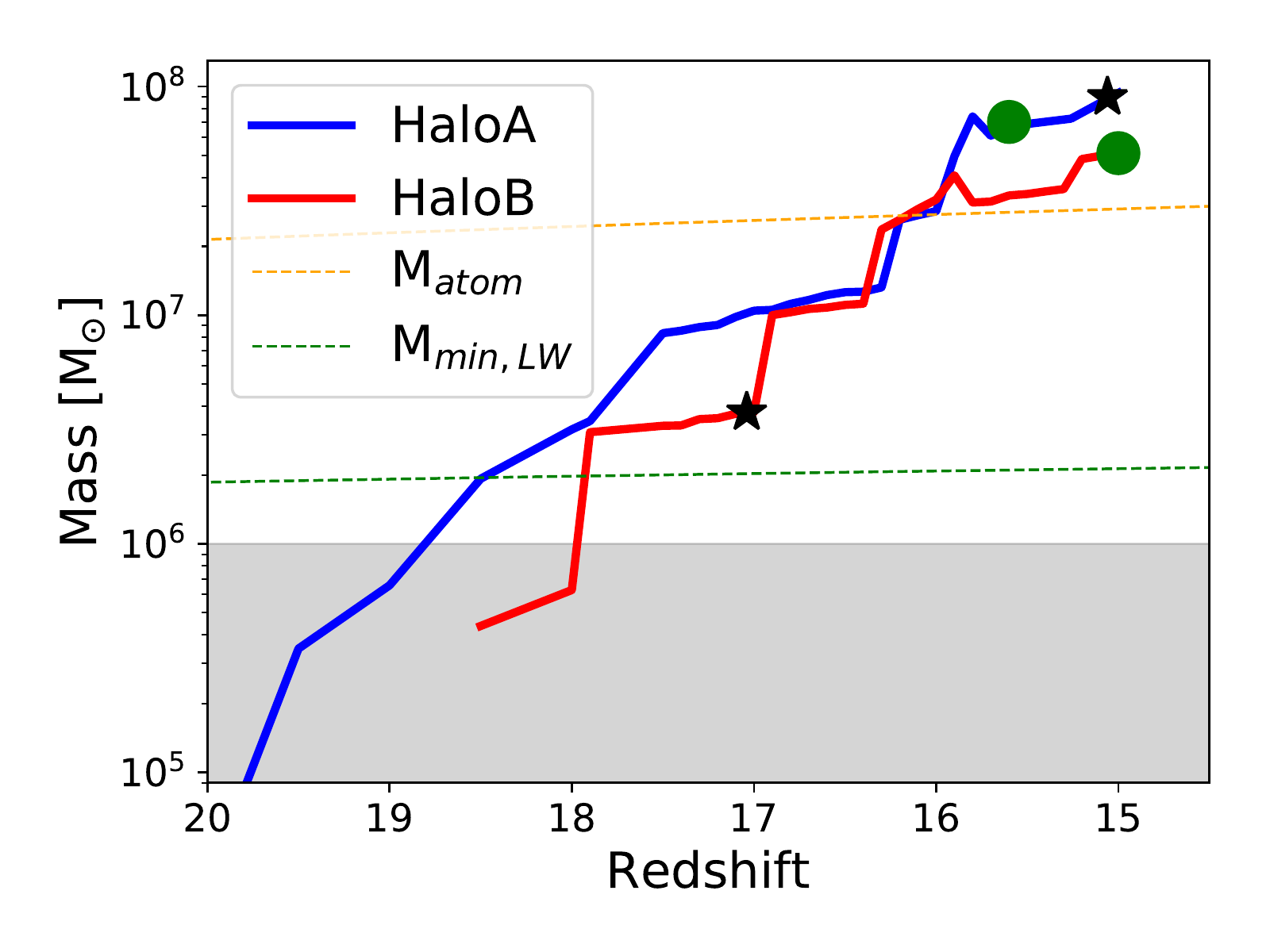}}
\caption{\textit{Left Panel:} The LW background field imposed on Halo A for the re-simulation.
  \hb is not shown as no LW background is imposed on \hbc. The LW field values are composed of local
  sources, plus a LW background from \citet{Wise_2012b}, extracted from the original Renaissance
  simulations. The local sources dominate over the background by at least an order of magnitude at
  all times. In the original Renaissance simulations \ha (and \texttt{HaloB})  remained metal-free
  and star-free until $z = 15.6$ ($z = 15.0$). The redshift of original collapse is marked with a
  green circle. A black star marks the redshift at which star formation occurs in \ha in the
  zoom-in simulations.
  \textit{Right Panel:} The merger history of both \ha and \hb as determined from the original
  Renaissance simulations. Again we mark the redshift of first star formation
  in \ha and \hb as found in the re-simulations with black stars. Note that star formation now
  occurs significantly earlier in \hb because no LW background is imposed. In \hac, on the other
  hand, star formation occurs 15 Myr after the halo was detected in the original (lower resolution)
  simulations. M$_{\rm atom}$, orange line, is the atomic cooling threshold \citep{Fernandez_2014}
  and M$_{\rm min,LW}$ is the threshold mass for halo collapse in the presence of an expected LW
  background at these redshifts \citep{Machacek_2001, OShea_2008}.
  }  \label{Fig:LWHistory}
\end{center} \end{minipage}

\end{figure*}

 \subsection{Re-simulation of pristine atomic cooling haloes} \label{Sec:Resimulation}
In order to make the simulation time tractable, we restricted the mesh refinement to be focused
around the target haloes. In order to do this we first identified the Lagrangian
volume (setting this to three times the virial radius) of the target halo
at the redshift at which it was first identified as a star-free, atomic cooling halo with zero
metallicity ($z = 15.6$, M$_{halo} \sim 7 \times 10^7 $\msolar for \ha and $z = 15.0$,
M$_{halo} \sim 5 \times 10^7$ \msolar for \hbc) and tagged each dark matter particle within this volume. This ensured
 that we  captured the dynamics of the gas and dark matter both within and surrounding the halo.
 Upon restarting
 the simulation subsequently at $z = 20$ we then set each of the tagged dark matter particles as
 ``must-refine-particles.'' This meant that any cell, in the simulation, containing one of these
 particles was allowed to refine. Any cells not containing one of these particles could not refine.
 This optimisation focuses the refinement solely onto the target halo (and gas surrounding the halo).
 In addition to the ``must-refine-particle'' refinement criteria,  refinement is updated to be
 also based on the Jeans length of the
 gas with the additional criteria that the Jeans length is always refined by at least 64 cells. \\ 
 \indent Having now optimised the simulations to focus only on the target halo 
 we next looked to (dark matter) particle splitting. By splitting the dark matter
 particles we increased the mass resolution of the simulation so as to match the increased spatial
 resolution. Dark matter particles are split in \enzo using the prescription of
 \cite{Kitsionas_2002} and was previously described in \cite{Regan_2015}. By employing particle
 splitting we reduced the dark matter particle mass to $M_{\rm DM} \sim 170$ \msolar inside the
 target halo and its surrounding Lagrangian volume. 

 \subsection{The External Radiation Field} \label{Sec:LWRadField}
 With refinement targeted only on a single halo (and its progenitors) star formation in the
 surrounding galaxies is therefore neglected. In order to account for the radiation that would
 otherwise be emitted by these galaxies we extracted the LW emission from the original simulations
 and created a table of LW values that this target halo is exposed to as a function of
 redshift. By doing this the spatial information of the LW field is preserved and so the target haloes feel the same LW radiation as they did in the original simulations where star formation and feedback was followed in every galaxy in the volume. Using the \grackle \citep{Grackle} software library we then used these LW tables as
 the ``background'' that this halo is exposed to. Self-shielding to LW radiation is also invoked in
 the re-simulations based on the \cite{Wolcott-Green_2011} prescription.\\
 \indent For the re-simulation of \ha the LW radiation background follows exactly the radiation
 field as determined from the original simulations and is shown in Figure \ref{Fig:LWHistory}.
 For \hb we purposely set the background radiation field to zero. This allows us to study the
 evolution of \hb with star formation suppression due to dynamical heating only. \hb is therefore
 the control simulation\footnote{Ideally \ha without a LW field would have been the
     control for
   the simulation campaign. However, as noted in footnote 4 this was not possible at this
   time.} and allows us to examine the impact of what happens when
 no LW field is present. 
 
 \subsection{Subgrid Star Formation Prescription} \label{Sec:StarFormation}
 \indent In order to resolve star formation in the collapsing target haloes we set the maximum refinement
 level of the simulation to 20. This is an increase of a factor of $2^8 (256)$ compared to the
 original Renaissance simulations and allows us to reach a maximum spatial resolution of $\Delta x \sim 1000$~au.
 While this (maximum) resolution is less than what was achieved
 in \cite{Wise_2019} it was necessary as the goal of this re-simulation was not only to follow the
 collapse of the target halo but to also follow the formation of stars within the collapsing
 halo for up to 2 Myr following the formation of the first star. At the resolution used in
 \cite{Wise_2019} this proved intractable and so we reduced the
 resolution by a factor of $2^4 (16)$, compared to \cite{Wise_2019}, as a compromise. Reducing the
 refinement factor compared to \cite{Wise_2019} reduced the computational load while still allowing us to
 resolve star formation at an acceptable resolution. \\
 \indent In order to model star formation within the collapsing gas cloud we employed a star
 formation criteria using the methodology first described in \cite{Krumholz_2004}. The implementation
 in \enzo is described in detail in \cite{Regan_2018a} and \cite{Regan_2018b} and we give a
 brief overview here for completeness. Stars are formed when all of the following conditions are met:
\begin{enumerate}
\item The cell is at the highest refinement level
\item The cell exceeds the Jeans density 
\item The flow around the cell is converging 
\item The cooling time of the cell is less than the freefall time
\item The cell is at a local minimum of the gravitational potential
\end{enumerate}
Once the star is formed accretion onto the star is determined by evaluating the mass flux across a
sphere with a radius of 4 cells centered on the star. Initially all stars are assumed to be stars with low surface
temperatures that are appropriate for main sequence SMSs and less massive proto-stars on the
Hayashi track. The accretion onto the surface of the embryonic star is found by applying Gauss's
divergence theorem to the volume integral of the accretion zone \citep[e.g][]{Bleuler_2014}
(i.e. the volume integral of flux inside the accretion zone)
\begin{equation}
  \dot{M} = 4\pi \int_\Omega { \rho v_r^- r^2 dr}
\end{equation}
where $\dot{M}$ is the mass accretion rate, $\Omega$ is the accretion zone over which we integrate,
$\rho$ is the
density of the cells intersecting the surface, $v_r^-$ is the velocity of cells intersecting
the surface which have negative radial velocities and $r$ is the radius of our surface. As noted above we
set the accretion radius to be 4 cells. The accretion onto the star is calculated at each timestep,
however this is likely to be a very noisy metric. To alleviate this to some degree we average
the accretion rate over intervals of 1 kyr and use that averaged accretion rate in data outputs.
The accretion rate is added as an attribute to each star and hence a full
accretion history of every star is outputted as part of every snapshot. Mergers with other stars
are also included in the accretion onto the stars. Simulations of massive star formation
\citep[e.g.][]{Meyer_2020} do show that stellar mergers do increase the “bloatedness” of a star
and hence including the mergers in the accretion rates is likely to be valid.
In this case the more massive star retains its
information (e.g., age, type, etc.) after the merger event - information on the less massive star is
lost. The mass of the less massive star is added to the accretion rate of the more massive star for
that timestep. Stars are merged when they come within 3 times the accretion radius of each other (i.e. 12 cell lengths).
We also note here that the dark matter particle mass is approximately $M_{\rm DM} \sim 170$ \msolarc.
  This mass is comparable to and greater than the mass of some of the stars that are formed. While
  the star forming regions are strongly baryon dominated individual dark matter particles may produce
  some shot noise. The shot noise effects are accounted for however by smoothing the dark matter particles
  over a scale of approximately 1 pc. This alleviates the impact of individual dark matter particles
  on the central gas and star particles \citep[e.g.][]{Regan_2015}. \\
\begin{figure*}
\centering
\begin{minipage}{175mm}      \begin{center} 
\centerline{
\includegraphics[width=0.49\textwidth]{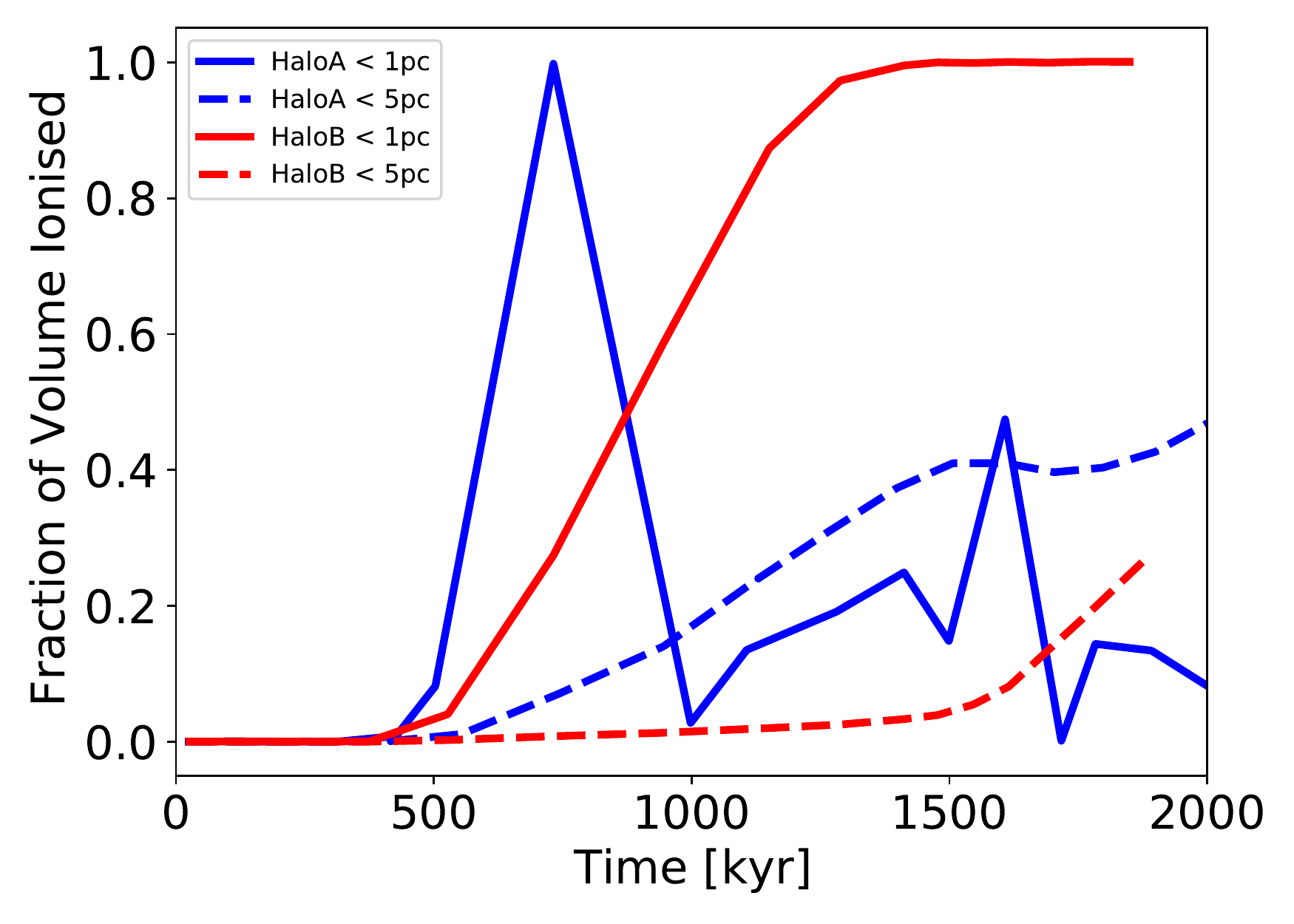}
\includegraphics[width=0.49\textwidth]{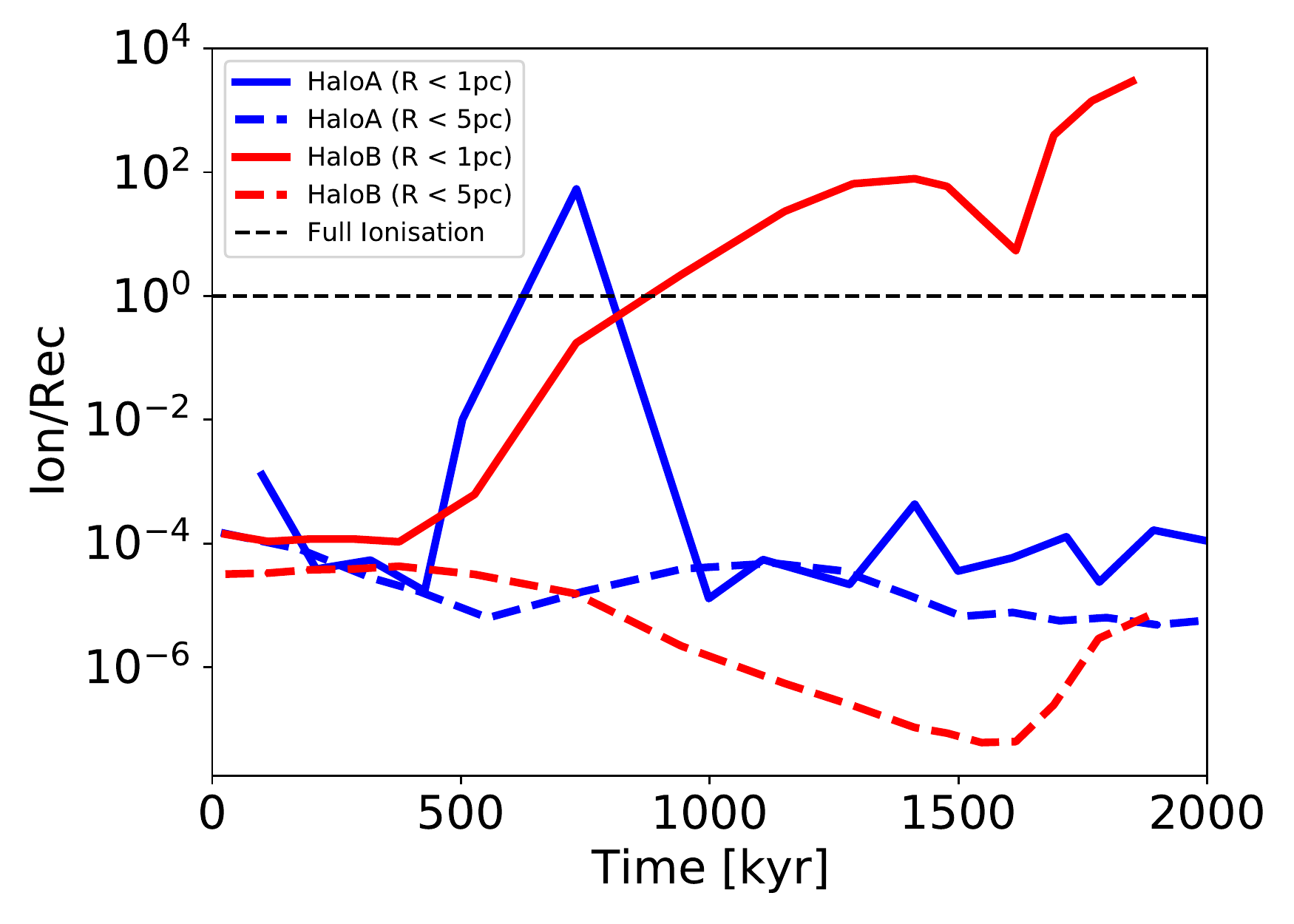}}
\caption{\textit{Left Panel:} The percentage of the inner volume ionised in each halo
  due to all stars as a function of time. In each case the volume is centred on the
  most massive star at that time. Two radii are considered, a radius of 1 pc from the
  most massive star is shown as a solid line, a radius of 5 pc is shown as a
  dashed line. \ha is shown in blue, \hb in red. For both haloes the 1 pc volume surrounding the
  most massive star becomes ionised but as we look to the 
  larger region (5 pc) the medium remains neutral. Note that for \ha the
  most massive star changes over time and this is reflected in the ionisation spikes.
  The first spike, where the volume becomes ionised quickly (at approximately 700 kyr) is at
  least partially due to the star moving to a lower density environment and hence ionisation
  becomes easier. 
  For \hbc, after approximately 1800 kyr the gas within 1 pc of the most massive star has become
  fully ionised. However, at a radius of 5 pc the gas shows virtually no ionisation. So while the
  stars are able to ionise their immediate surroundings
  the radiative feedback does not penetrate much further than a pc. 
  \textit{Right Panel:} The ratio of the ionising photon luminosity and the total integrated
  recombination rate of the gas. For \ha we see a similar trend to the left panel. The most massive
  star, at a time less than $\sim 750$ kyr is quickly able to ionise its surroundings. However, at
  larger radii (and also at later times for \ha when the most massive star is embedded in the
  central region) the gas remains neutral. 
  For \hb the results are again similar to the volume averaged
  calculation. At 1 pc the region becomes fully ionised after approximately 1000 kyr but again
  at a radius 5 pc the region remains dominated by recombinations.
  }  \label{Fig:Ionisation}
\end{center} \end{minipage}

\end{figure*}
\indent Each star also has the ability to provide both radiative and mechanical feedback, which is
most appropriate in the case where the star has transitioned into a (accreting) black hole. As the
accretion rate onto the star varies the star can transition its type from a SMS, with an inflated
surface, to a PopIII star.  This transitioning only occurs if the accretion rate onto the star either
never exceeds the critical rate (set in our simulations be be 0.04 \msolaryrc) or if the accretion
rate onto the star falls below the critical accretion rate. While the star remains bloated the
radiative
feedback from the star is primarily below the hydrogen ionisation limit and is mostly in the
form of infrared radiation. However, if the accretion rate drops and the star contracts to the
main sequence then its surface temperature dramatically increases up to $10^5$~K,
causing its spectrum to harden and peak in the UV. \\
\indent All stars in this simulation emit radiative feedback below the ionisation threshold of
hydrogen. The radiation is followed explicitly using the ray tracing technique \citep{WiseAbel_2011}.
Pop III stars are modelled assuming a blackbody spectrum with a characteristic mass of 40 \msolarc
 (Table 4, \cite{Schaerer_2002}). From that we assign a \textit{LuminosityPerSolarMass} to the
Pop III star and the star consequently becomes more luminous and the feedback more intense as the mass
of the star increases.  SMSs are modelled by assuming
a blackbody spectrum with an effective temperature of T$_{\rm eff}$ = 5500 K \citep{Hosokawa_2013}.
The radiation spectrum for a SMS therefore peaks in the infrared as opposed to the UV for Pop
III stars. For the specific luminosity of the SMS we take a characteristic mass of 500 \msolar and
apply the contribution from the non-ionising photons only \citep{Schaerer_2002}. As with the 'normal'
Pop III stars the SMS luminosity changes as mass is accreted and the total luminosity then
scales up as the mass increases. \\
\indent In both cases the radiation from the stars is propagated outwards from the star using the
\texttt{MORAY} radiative transfer package \citep{WiseAbel_2011} that is part of \enzoc.
\texttt{MORAY} is able to model the ionisation of H, He and He$^{+}$. It can also account for the
photo-dissociation of \molH for photons with energies within the Lyman-Werner band and the
photo-detachment of  $\rm{H^-}$ and $\rm{H_2^+}$ for photons in the infrared band. For each type of
star we use five energy bins. The first two energy bins (E $< 13.6$ eV) are weighted by the cross
section peaks for $\rm{H^-}$,  $\rm{H_2^+}$ and \molH photo detachment/dissociation respectively.
The next three energy bins are determined using the \texttt{sedop} code developed by
\cite{Mirocha_2012} which determines the optimum number of energy bins needed to
accurately model radiation with energy above the ionisation threshold of hydrogen. For the
self-shielding of \molH against LW radiation we use the prescription of \cite{Wolcott-Green_2011}.\\
\indent As stars in this simulation contract to the main sequence, when the accretion rate onto the
star drops, radiation above the hydrogen ionisation threshold is not initiated although in
principle it should be and our radiative transfer scheme does support ionising radiation. The
problem however is that the shocks generated near
the star particle are too strong and unresolved for the PPM reconstruction scheme and HLLC Riemann
solver to handle, the steep gradients cause the solution to overshoot to negative densities and
energies \citep[see][for details]{Enzo_2014}. We could have avoided this problem by utilizing a more
diffusive solver at the expense of accuracy, but instead
we neglect the impact of ionising feedback and post-process the impact of
ionising feedback with \cloudy and use the results to estimate the extent to which the stars
(particularly the hyper luminous stars in \hac) ionise their surroundings and potentially shutdown
further star formation\footnote{Initially our simulation plan had been to run four haloes with
  and without a background LW field at a refinement level of 24 (16 times higher than we report
  here). However, the simulations proved intractable and computationally expensive and hence we
  reduced the scope of the campaign. As a result, we did not have the capacity to run
    \ha without a LW radiation field or \hb with a radiation field.
  We are now in the process of undertaking a new
  more expansive campaign to address these shortcomings.}.
 
\begin{figure*} 
\centering
\begin{minipage}{175mm}      \begin{center} 
\centerline{
\includegraphics[width=0.52\textwidth]{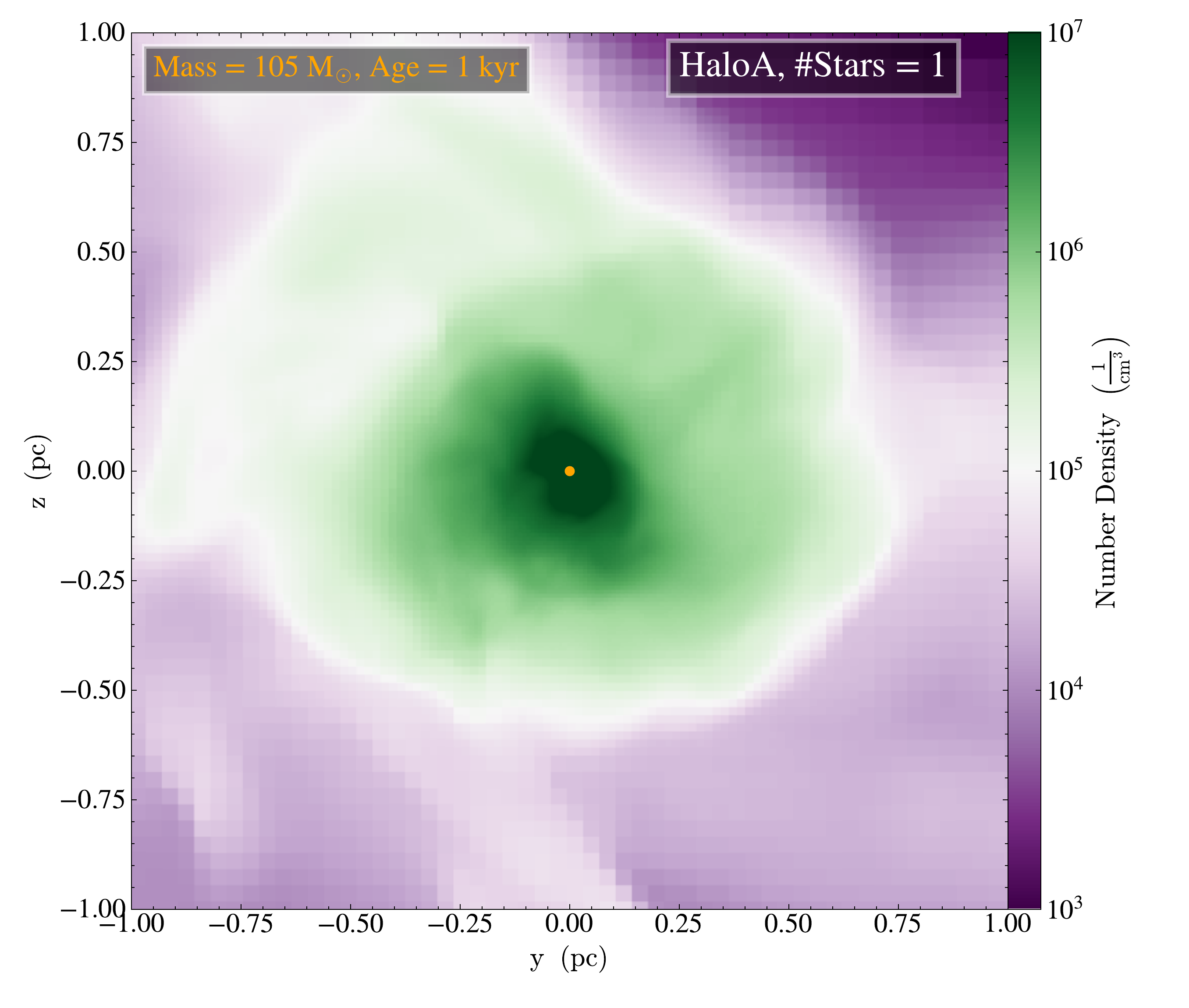}
\includegraphics[width=0.52\textwidth]{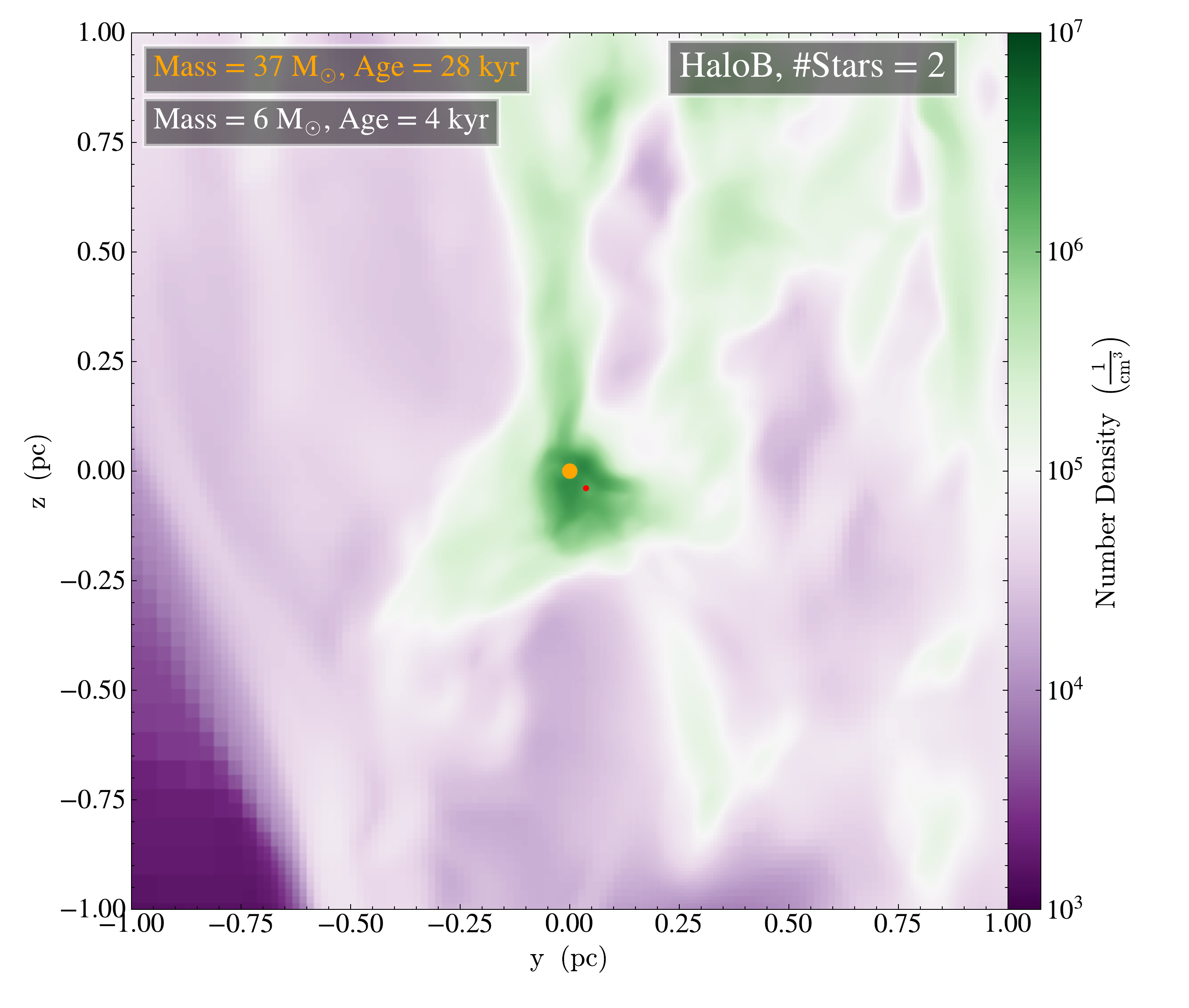}}
\caption{Both panels show the projected number density$^6$ of the region around which the first
  star forms. \ha is in the left hand panel, \hb in the right hand panel.
  The legend in each panel gives the mass of each star at the first output time following star
  formation as well as the age of each star at that time. The extent of each panel is 2 pc (physical). The orange
  star in each case represents the most massive star. Stars coloured in red are stars in which the accretion rate
  exceeds that required for supermassive star formation (0.04 \msolaryrc in our simulations).
  Blue stars are those stars
  (normal PopIII stars) for
  which the rate is below the critical rate. \ha contains a single initial star. \hb contains two stars (closely separated)
  at the first output following star formation. For illustrative purposes the size of each star in the
projection is scaled as $\rm{R_{star}} \propto \rm{M_{star}^{0.6}}$.}\label{Fig:ProjectionStart}
\end{center} \end{minipage}

\end{figure*}

\subsection{Post-Processing with Cloudy}
\label{cloudy:description}
      {\sc Cloudy} \citep{Ferland_2017} is a spectral synthesis code which models radiative transfer
      through a gas, and its resulting thermal and chemical equilibrium, under a wide range of conditions
      encompassing those expected for interstellar matter. 
      The code relies on a number of databases for computing the behaviour of atoms and molecules,
      including tabulated recombination coefficients obtained from \cite{Badnell_2003}
      and \cite{Badnell_2006}, with Case A and B recombination predictions for single-electron
      systems from \cite{Storey_1995} and He I recombination rates from \cite{Porter_2012},
      as well as ionic emission data from the CHIANTI database \citep{Dere_1997,Dere_2012}.
      Its chief limitation is that it is largely constrained to modelling environments in 1-D
      and/or assuming spherical symmetry \citep[though see, e.g.,][for recent efforts to extend
        its implementation to pseudo-3D problems]{Morisset_2013, Fitzgerald_2020}. For this
      reason, and because we do not follow the stellar evolution of each formed star in detail,
      we must make a number of simplifying approximations in using {\sc cloudy} to estimate
      the impact of ionising feedback. 

For each halo, we consider the ionising feedback for a range of times between 1~kyr and 2000~kyr
after the initial star formation has commenced. For each snapshot, we divide the stars formed between
those which are accreting $> 5 \times 10^{-3}$ \msolaryrc, which will evolve on the Hayashi track due
to $\rm{H}^{-}$ opacity in their inflated envelopes \citep[e.g.,][]{Hosokawa_2013}, and those accreting
below this threshold, which will evolve blue-ward as they contract to become hot ionising sources on
the ZAMS \citep{Haemmerle_2017}. We assume that all rapidly-accreting (``red'') stars remain negligible
ionising sources, and that all slowly- or non-accreting (``blue'') stars have thermally-relaxed to a
main sequence temperature of $\approx 10^{5}$K \citep{Schaerer_2002, Woods_2020}. We further take
their luminosities to be approximately Eddington (L $\approx 1.3\times 10^{38} \times (M/M_{\odot})$
erg s$^{-1}$), and their spectra to be well-approximated as blackbodies.

For each blue star, we then model the ionisation state of the surrounding gas assuming spherical
symmetry, with the density profile found from the average gas density in successive shells 0.5pc
in width, centred on each star in the halo, and primordial abundance ratios taken to be
1.0:0.08232:1.6e-10:1e-16 for H:He:Li:Be \citep[consistent with the results of the][table 2, see
  {\sc cloudy} documentation for further discussion]{Planck_2014}. We assume an inner boundary
of $10^{15}$ cm ($\sim 3 \times 10^{-4}$ pc) and terminate our calculations at the outer boundary of
each nebula once either the gas temperature falls to 8000 K or the electron fraction falls below 5\%.
Modelling the ionised nebula associated with each star in this way does not account for the
overlapping of Str{\"o}mgren spheres associated with distinct stars; we address this point and
further limitations in $\S$\ref{cloudy:results}.
\begin{figure*}
\centering
\begin{minipage}{175mm}      \begin{center} 
\centerline{
\includegraphics[width=0.52\textwidth]{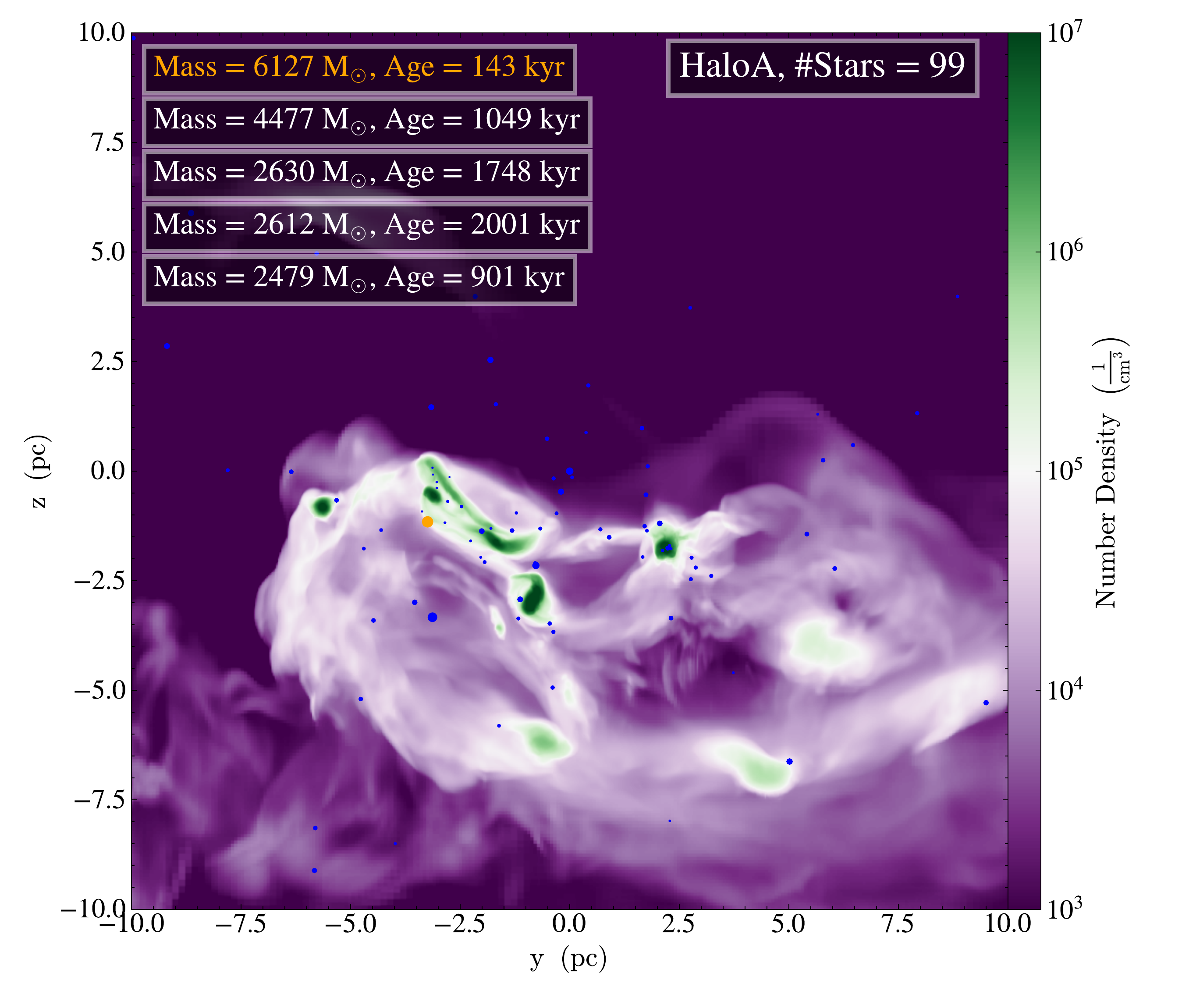}
\includegraphics[width=0.52\textwidth]{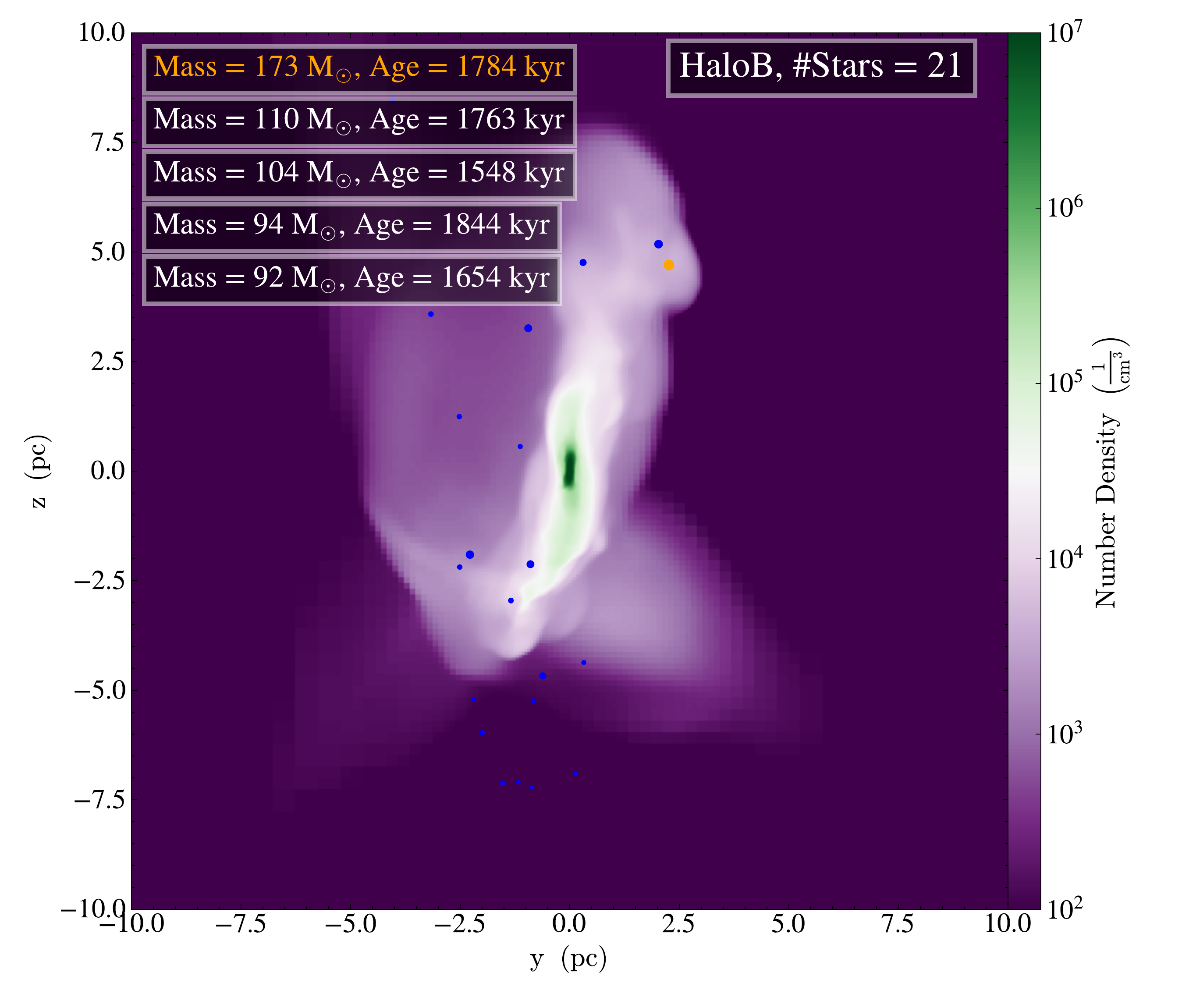}}
\caption{Both panels show the projected number density$^6$ in each halo at the end of each
  simulation. The left panel shows \ha while the right panel shows \hbc.
  The legend in each figure gives the mass of the five most massive stars at the final output
  time, as well as
  the age of the star at that time. The extent of each panel is 20 pc (physical). The orange
  star in each case represents the most massive star. Blue stars are those stars
  (normal PopIII stars) for which the rate is below the critical rate (0.04 \msolaryrc) for
  SMS formation. At the times shown no
  SMSs exist in either simulation because the accretion rate onto each star is less than the
  critical rate.  For illustrative purposes the size of each star in the
  projection is scaled as $\rm{R_{star}} \propto \rm{M_{star}^{0.6}}$.}  \label{Fig:ProjectionEnd}
\end{center} \end{minipage}

\end{figure*}

\section{Results} \label{Sec:Results}

\subsection{Conditions for (Super-)Massive Star Formation}
As noted in \S \ref{Sec:Introduction} the goal of this study is to model the formation of
SMS formation in haloes which are experiencing both LW feedback from nearby galaxies and
dynamical heating from mergers. Both of these effects suppress the formation of regular
PopIII stars in haloes
below the atomic cooling limit and hence may lead to the formation of a SMS in a larger
halo  \citep[as was found in][]{Wise_2019}. In this study we include 
sub-grid star formation prescriptions to follow the star formation process in 
two of these candidate haloes.\\
\indent In the left hand panel of Figure \ref{Fig:LWHistory} we show the LW history that \ha is
exposed to throughout its re-simulation. As discussed in \S \ref{Sec:Methods} the LW field at the
location of the halo is determined from the original simulation, which included star formation
in all of the surrounding haloes. In the re-simulation the adaptive mesh refinement
grids are focused only around the
target halo and so the LW field must be imposed as a global background, albeit focused on a single
halo. As can be seen in the left hand panel of Figure \ref{Fig:LWHistory} the LW field is
very flat and sits below \JLW \ $\sim 2$
J$_{21}$ until a redshift of $z \sim 16.5$ at which point it increases strongly due to a nearby source
which undergoes a starburst and emits copious amounts of LW radiation
(and ionising radiation\footnote{The
  shorter mean free path of the ionising radiation means that \ha does not get photoionised.}) in the
direction of \hac. The LW field reaches its zenith at a redshift of $z \sim 15.8$ and thereafter
starts to decline. The green circle at $z = 15.6$ represents the redshift at which the halo was
detected as a metal-free and star-free atomic cooling halo in the original simulations
\citep[see][for details]{Regan_2020}. The black star indicates the redshift at which star
formation begins in this re-simulation ($z = 15.05$). \\
\indent In the right hand panel we show the merger history of both \ha and \hb
as found in the original simulations. The mergers, both major and minor, drive the dynamical heating
which heats the gas and delays star formation \citep{Wise_2019}. 
\ha experiences a major merger starting at $z \sim 16.2$ and lasts until $z \sim 15.7$. This
merger, in combination with the LW field, suppresses star formation until z = 15.05 (again the green
circles denotes the time of collapse in the original, lower resolution, simulations). The merger
history of \hb is shown as the red line. Recall that \hb is run without any LW background as
a control case. As a
result star formation is triggered at $z = 17.21$. Using the results from the original simulation
which allows us to see the future evolution of that halo we see that immediately after star formation
the halo undergoes a major merger. It is therefore highly likely that, in the absence of a suppressing
LW background, the merger triggers star formation, which in the original simulation was
suppressed until after $z = 15$. Therefore, we can see here that the absence of the background
LW radiation field was sufficient to allow PopIII star formation to take place. Dynamical heating
by itself was not sufficient for this halo to avoid star formation. \\
\indent Star formation occurs in both \ha and \hb when they are at very different phases in
their evolution.
In \ha the mass of the halo is significantly above the atomic cooling limit
(M$_{\rm{HaloA}} = 9.3 \times 10^7$ \msolarc) and the halo has
remained metal-free. \hbc, on the other hand, collapses early with a mass of
M$_{\rm{HaloB}} = 3.7 \times 10^6$ \msolar due to the lack of a LW background and thus can
be classified as a mini-halo. Before analysing the star formation in detail we use results
from \cloudy to determine if the negative radiative feedback from stars can
quench further star formation through photoionisation in each halo.

\subsection{Determining the time of star formation quenching}\label{cloudy:results}
To assess the impact of photoionising stellar feedback, we must investigate the ionisation state
of the gas in the innermost star-forming regions of each halo.
Here we adopt two different radii at which to calculate the ionisation state of the gas.
We use a radius of 1 pc to estimate the ionisation state of the gas close to the most massive star
and we also use a radius of 5 pc to estimate the more ``large scale'' ionisation of the entire halo. 
We use the same radii, at which to estimate the ionisation state of the gas, for both haloes
for consistency. We then evaluate the evolution of both the total ionised volume and the
total ionisation budget through each snapshot, in order to provide subtly distinct but
complementary measures of the strength of stellar feedback.

\begin{figure*}
\centering
\begin{minipage}{175mm}      \begin{center}
\centerline{
    \includegraphics[width=18.0cm, height=12cm]{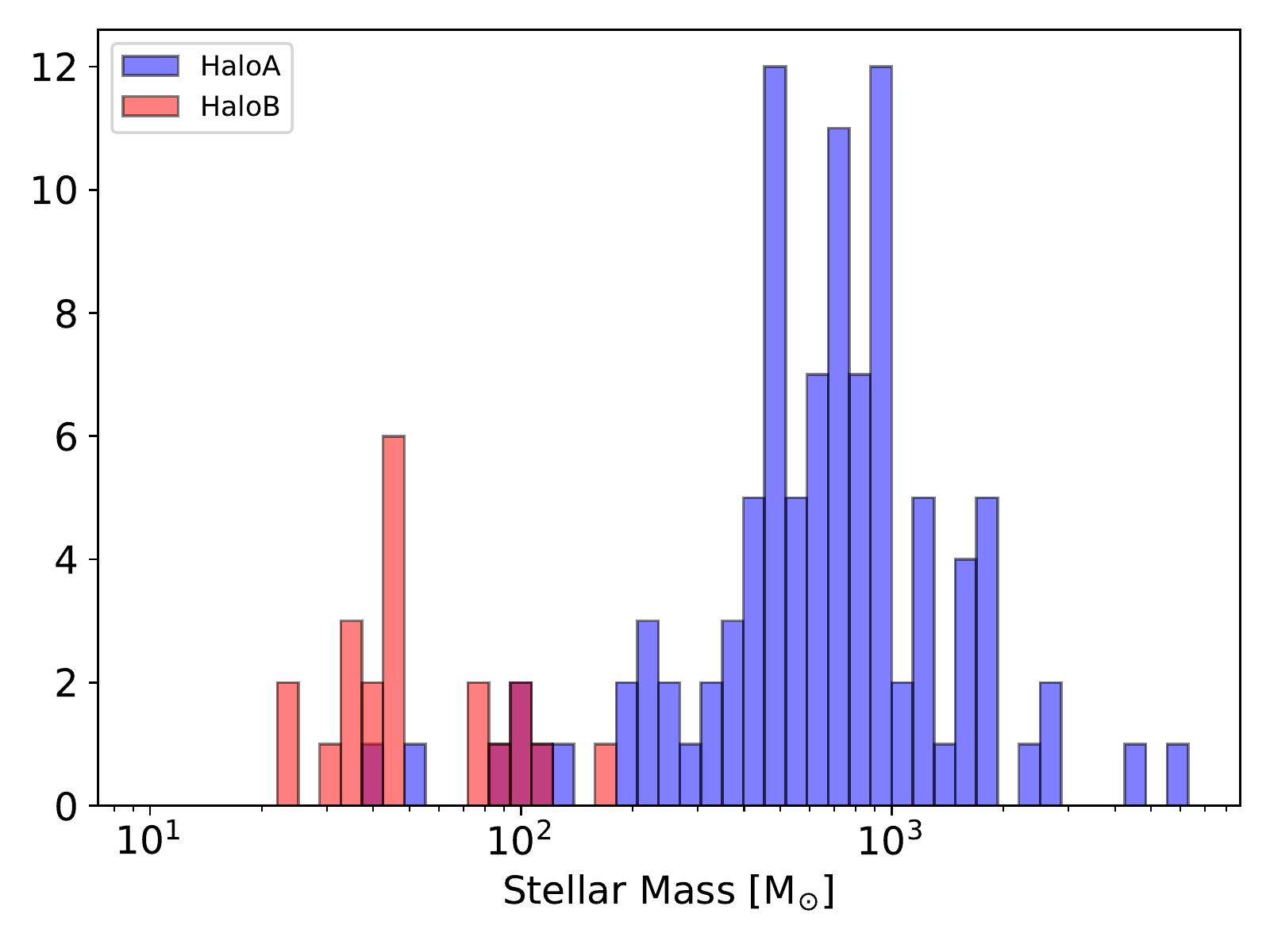}}
\caption{
  The mass function for the stars that formed in \ha \& \hb over approximately 2 Myr.
  Blue bars represent stellar masses from \ha while red bars represent stellar masses from \hbc.
  The dichotomy in masses is clearly evident with a strong bias towards more massive stars
  in \hac. This is due to the higher temperatures in \ha compared to \hb which in turn
  leads to high infall rates to the centre and hence more mass available for proto-stars
  to accrete. The median mass for a single star in \ha is 683 \msolar and the median mass
  for \hb is 44 \msolarc. 
}
\label{Fig:MassFunction}
\end{center} \end{minipage}
\end{figure*}
\begin{figure*}
\centering
\begin{minipage}{175mm}      \begin{center}
\centerline{
    \includegraphics[width=18.0cm, height=12cm]{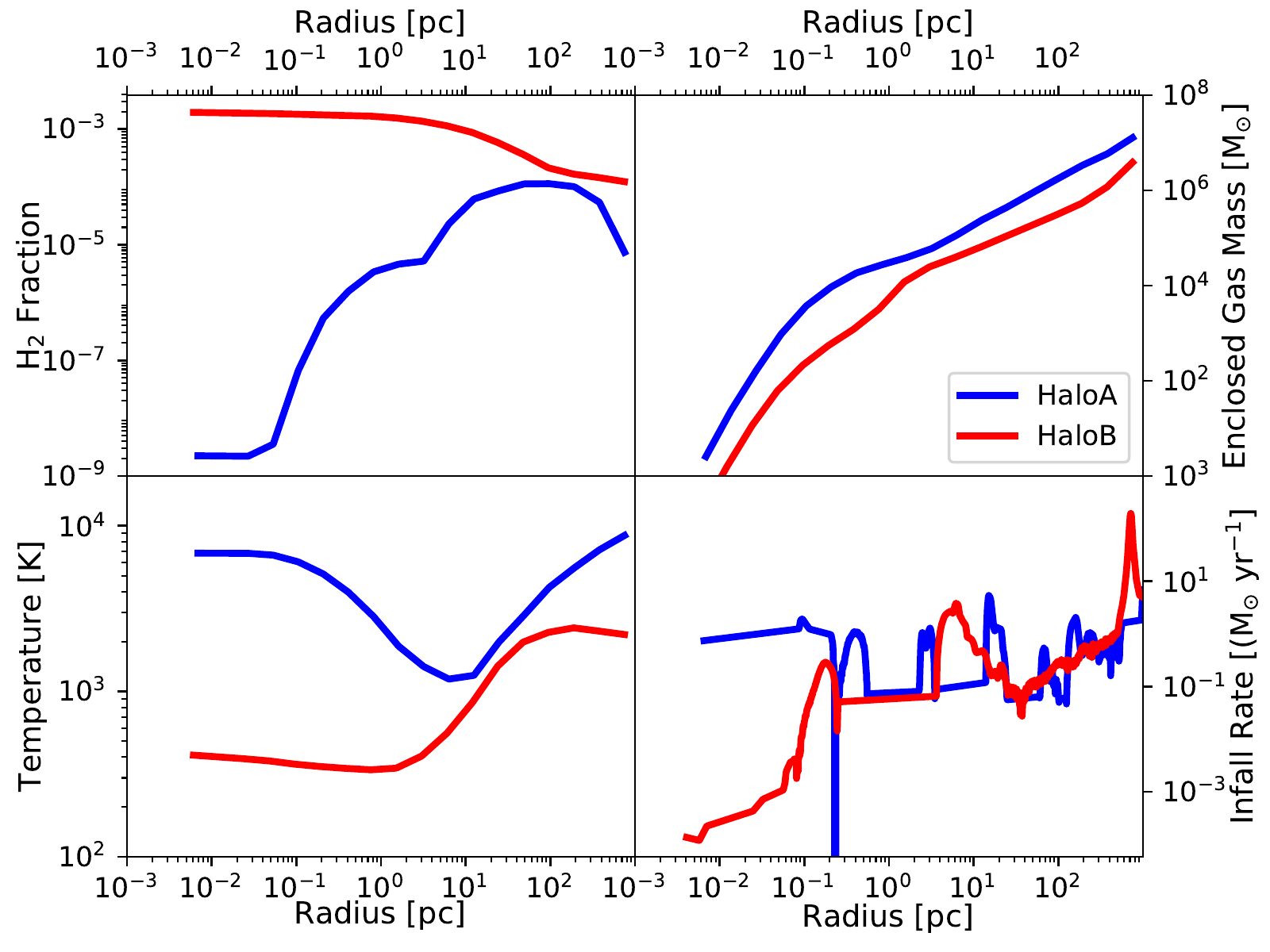}}
\caption{We examine the radial profiles for both \ha (blue line) and \hb (red line) immediately prior to star formation. The four panels
  are from bottom left going clockwise: temperature, \molH fraction, enclosed gas mass and infall rate.
  As \hb collapsed early due to the lack of a LW background (compared to the original Renaissance simulation)
  the temperature of this halo is significantly lower in the centre. The central temperature equilibrium
  value is close to 500 K which is characteristic of PopIII simulations for the chemical network used in this work.
  This lower temperature is due to the high \molH fraction, as shown in the upper left panel. This can
  be contrasted with the radial profiles for \ha which show systematically higher temperatures in the core of the
  halo and lower \molH fractions. The lower \molH fractions are due to a combination of dynamical heating from
  mergers and from local radiation sources. The enclosed mass fractions identify the difference
  in masses between the haloes. In the case of \hb the enclosed masses are systematically lower compared
  to \ha as this is a effectively a minihalo. Infall rates in both haloes are very different in the
  centre of each halo. Average infall rates in the very centre (R $\lesssim 0.1$ pc) of \ha are
  above 0.1 \msolaryr compared to below 0.001 \msolaryr in \hbc. These infall rates are consistent
  with the accretion rates that are seen onto the stars that subsequently form in each halo. The flat
  patches in the infall rate are due to outflow at those radii
  (which show up as negative infall rates). }
\label{Fig:RadialProfiles}
\end{center} \end{minipage}
\end{figure*}

First, we compute the ionised volume associated with each thermally-relaxed star as
described in \S\ref{cloudy:description}, and integrate the intersection of the union of these
Str{\"o}mgren regions within the central star-forming regions, as defined in the preceding paragraph
for each halo, using a Monte Carlo approach. Note that for a constant combined ionising luminosity,
any reduction in ionised volume in accounting for overlaps in our spherically-symmetric
Str{\"o}mgren spheres would in principle be compensated by the ionisation of a greater
(though presumably not spherically symmetric) volume by the remaining available
ionising photons. The extent to which this would raise the total ionised volume beyond
our estimate depends sensitively on the local density distribution beyond each Str{\" o}mgren region.

Indeed, an alternative approach is to simply compare the ionising photon luminosity of all stars
within the inner star-forming region with the total integrated recombination rate $\dot{\mathcal{N}}_{\rm R}$ of the gas if
all hydrogen atoms were ionised; the latter sets the total emission rate of photons with
$E>13.6$~eV needed to maintain ionisation of the star-forming region in equilibrium. This can be
found from integrating over the hydrogen number density:

\begin{equation}
  \dot{\mathcal{N}}_{\rm R} = \int_{\rm{V}}\alpha_{\rm{B}}(\rm{H}^{0},T)n_{\rm{H}}^{2}(r)d\rm{V}
\end{equation}

\noindent where $n_{\rm{H}}(r, t)$ is the spherically-averaged density profile centered on the
most massive star in the Halo at time $t$ (see above), and $\alpha_{\rm{B}}(\rm{H}^{0},T)$ is the
case B recombination coefficient for hydrogen, which we take
as $\approx1.4\times 10^{-13}\rm{cm}^{3}\rm{s}^{-1}$ for primordial gas at a temperature typical of
plasmas photoionised by Population III stars, $\rm{T}\sim 2\times 10^{4}$K
\citep{Osterbrock2006, Johnson2012}. 

These two measures of the strength of ionising stellar feedback are compared in
Figure \ref{Fig:Ionisation}. The left panel shows the volume ionised fraction of the innermost
region as a function of time, while the right panel shows
the fractional ionising budget as a function of time. The ionisation state of the gas is shown in
blue for \ha and red for \hbc. Calculations with a 1 pc sphere are shown as solid line, calculations
with a 5 pc radius are shown as dashed lines. Note that the ionised volume (left panel)
and the fractional ionising photon budget (right panel) do not necessarily
move perfectly in tandem. This is partly due to our simplified treatment as described above,
including overlapping Str{\" o}mgren regions and spherically-averaged density profiles, but also
reflects the variable density distribution, with the lowest density regions preferentially
ionised over high-density knots. \\
\indent We begin by focusing on \hac. In the left hand panel, we see a marked rise in the
ionised volume between $\sim 500$ and $\sim 1200$ kyr when measured in a sphere of 1 pc, this is
matched in both cases by a similar rise in the fractional ionising photon budget (right panel).
In \ha we see a sharp rise matched by a sharp drop (solid blue line). This is because the most
massive star in the simulation switches at approximately 1000 kyr as another star's accretion rate
gives it a larger mass. Prior to this time the most massive star easily ionises its immediate
surrounding and this is at least partly because the star also moves into a lower density region
(something similar occurs in \hbc). \\
\indent After 1000 kyr the most massive star in \ha becomes embedded in
dense filaments and its ability to ionise even its immediate surroundings is blunted (and hence the
ionisation fraction hovers between approximately 0.1 and 0.4). The
right hand panel shows a similar result for \ha with the most massive star, prior to 1000 kyr, easily
able to ionise
its immediate surroundings. However, after this point, with the most massive star more centrally
located, recombinations dominate. When considering the larger 5 pc radius (dashed blue line)
ionisation becomes much more difficult. In the left hand panel we see that the 5 pc region
is approximately 40\% ionised by 2000 kyr but if we look at the right hand panel we see that
recombinations are still overwhelmingly dominant. This is because \ha contains a plethora of dense knots and
filaments in which recombinations are dominant and into which ionising photons cannot penetrate. \\
\indent For \hb both measures, at 1 pc,  show a fully ionised medium
after approximately 1000 kyr. The most massive star has a mass of 173 \msolar and crucially
it moves to a lower
density region away from the central over-density. However, like \hac, using a radius of 5 pc shows
a medium whose ionisation state remains low reflecting the fact that while the stars are very
massive and easily able to ionise their immediate surroundings they are not yet numerous enough
to ionise the entire halo. \\
\indent In summary, ionising stellar emission is unlikely to shut down star formation in
either halo prior to the onset of chemical and radiative emission from the supernovae
explosions expected to occur for some of the stars \citep[e.g. see][]{Heger_2003}
approximately 2 Myr  after their formation. In
the case of \ha this is particularly important. \ha contains a number of hyper-luminous PopIII stars,
however, the much denser gas in the central region of \ha is robust against stellar feedback
allowing star formation to proceed in these dense pockets, 
albeit with the overall volume ionisation growing in importance as the simulation progresses
(see left panel of Figure \ref{Fig:Ionisation}). The results from our \cloudy models have
  very recently been confirmed by the 1-D analysis of \cite{Sakurai_2020} who investigated the
impact of ionising feedback from a growing protostar in a very similar environment. They found,
similar to us, that no significant HII region forms during the evolution of the star.


\begin{figure*}
\centering
\begin{minipage}{175mm}      \begin{center}
    \centerline{
      \includegraphics[width=9.0cm, height=6cm]{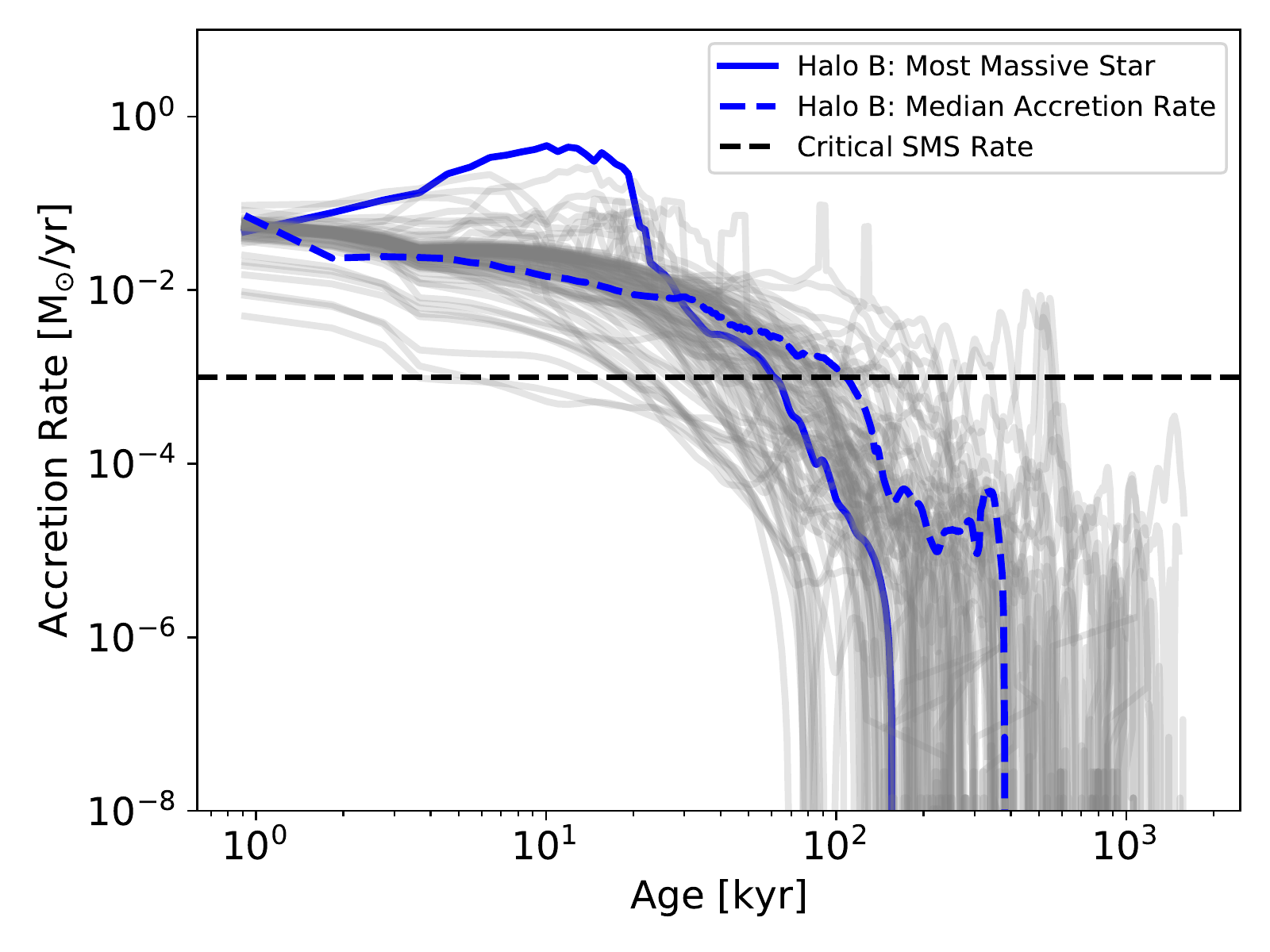}
      \includegraphics[width=9.0cm, height=6cm]{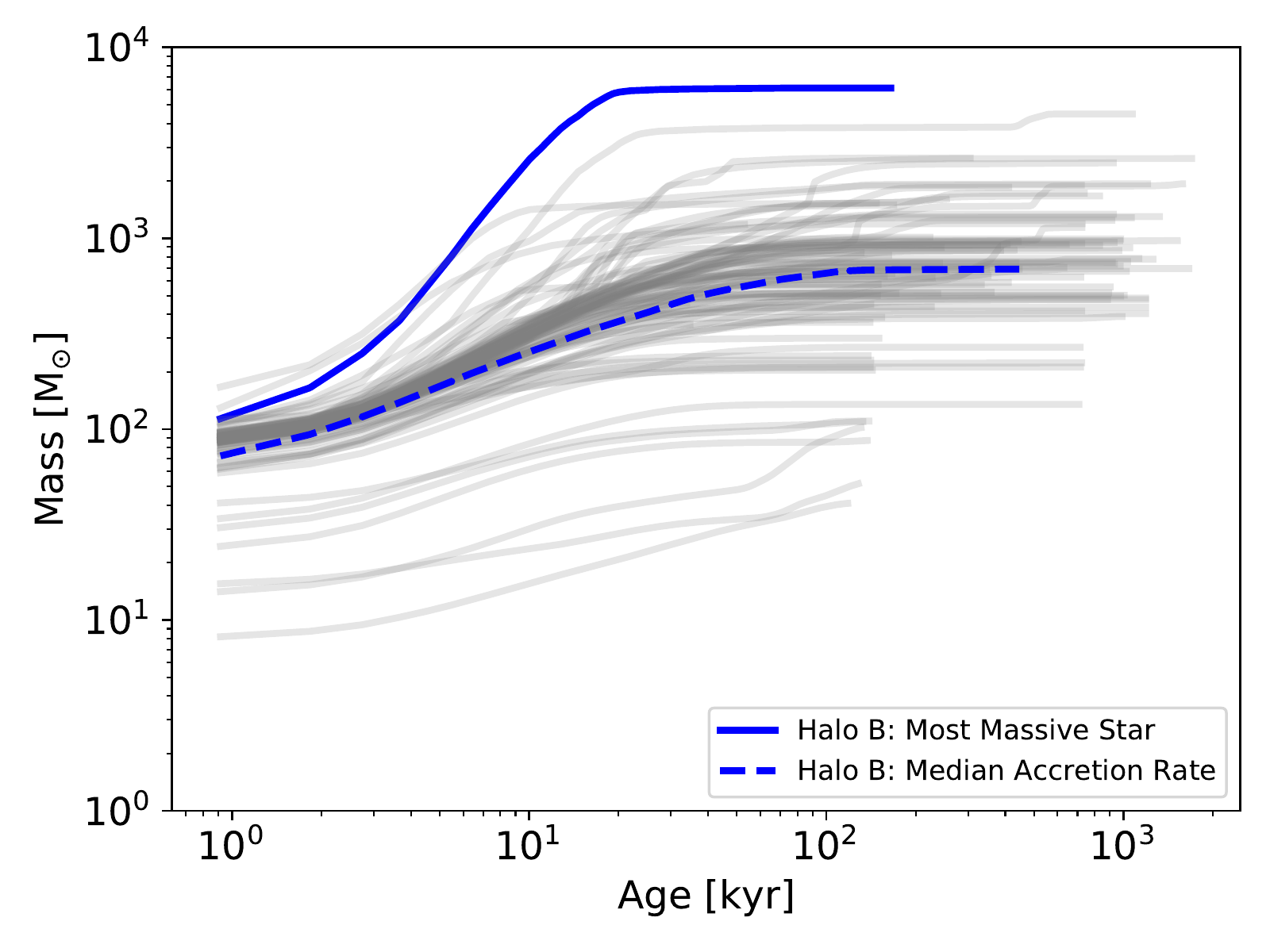}}
    \caption{
      \textit{Left Panel:} The accretion rate onto all stars in \hac. The mass accretion onto the
      most massive star is shown by the blue line. The dashed blue line gives the accretion
      rate onto the median mass star in \hac. Accretion onto all others stars are shown as grey lines.
      In all cases initial accretion rates decline over time, dropping below the critical rate
      for SMS formation after, on average, 100 kyr. The black dashed line gives the critical rate for SMS
      star formation \citep{Haemmerle_2017}.
      \textit{Right Panel:}  The mass evolution of every star \hac. The
      solid blue line is for the most massive star in \hac. It has a mass of
      6127 \msolar after 143 kyr. We also look at the mass evolution
      of an average star in \ha for comparison - given by the dashed blue line.
      In \ha a median$^4$ star has a mass of 683 \msolarc.
    }
\label{Fig:HaloA_AccretionRates}
\end{center} \end{minipage}
\end{figure*}



\begin{figure*}
\centering
\begin{minipage}{175mm}      \begin{center}
    \centerline{
      \includegraphics[width=9.0cm, height=6cm]{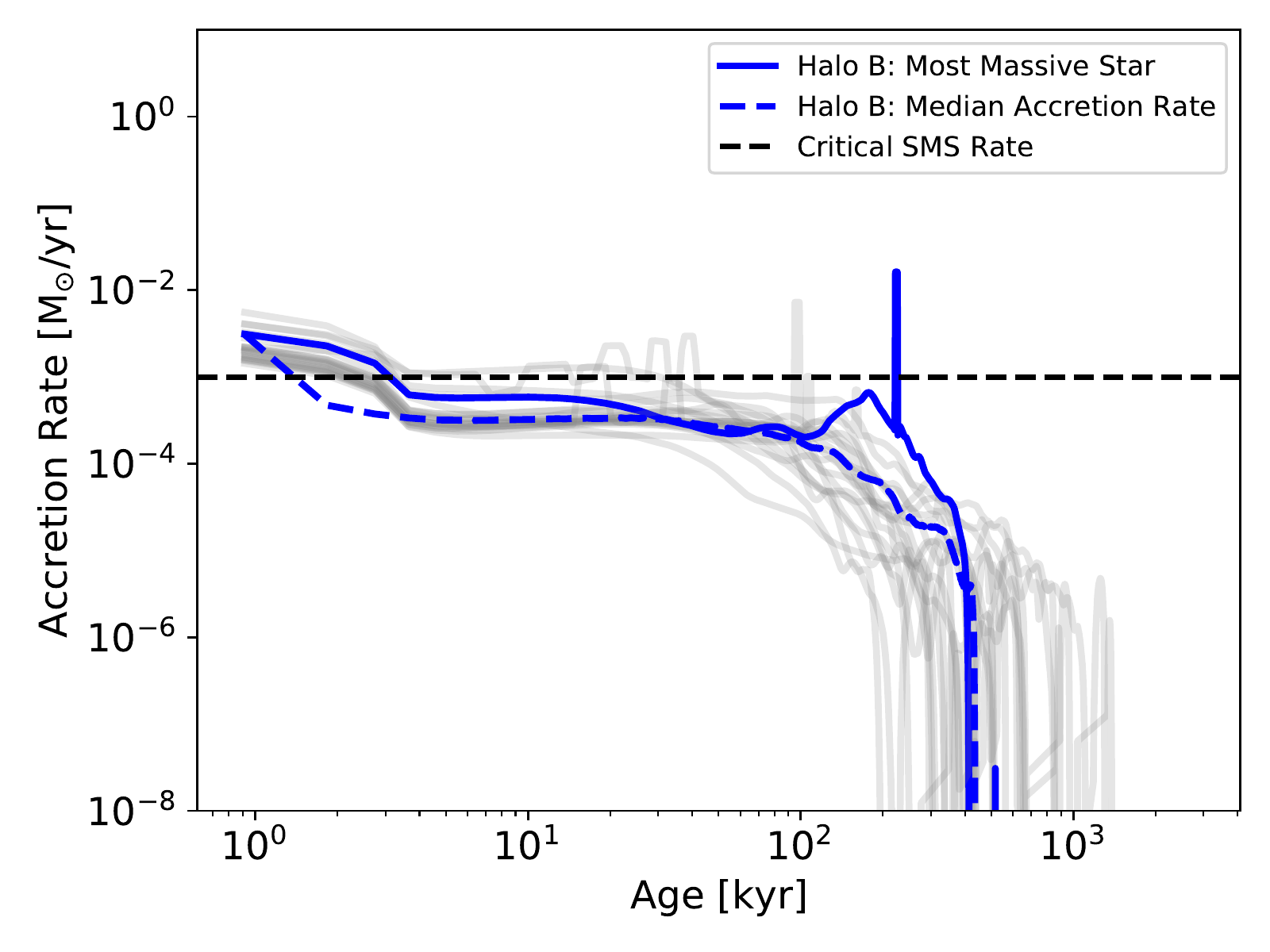}
      \includegraphics[width=9.0cm, height=6cm]{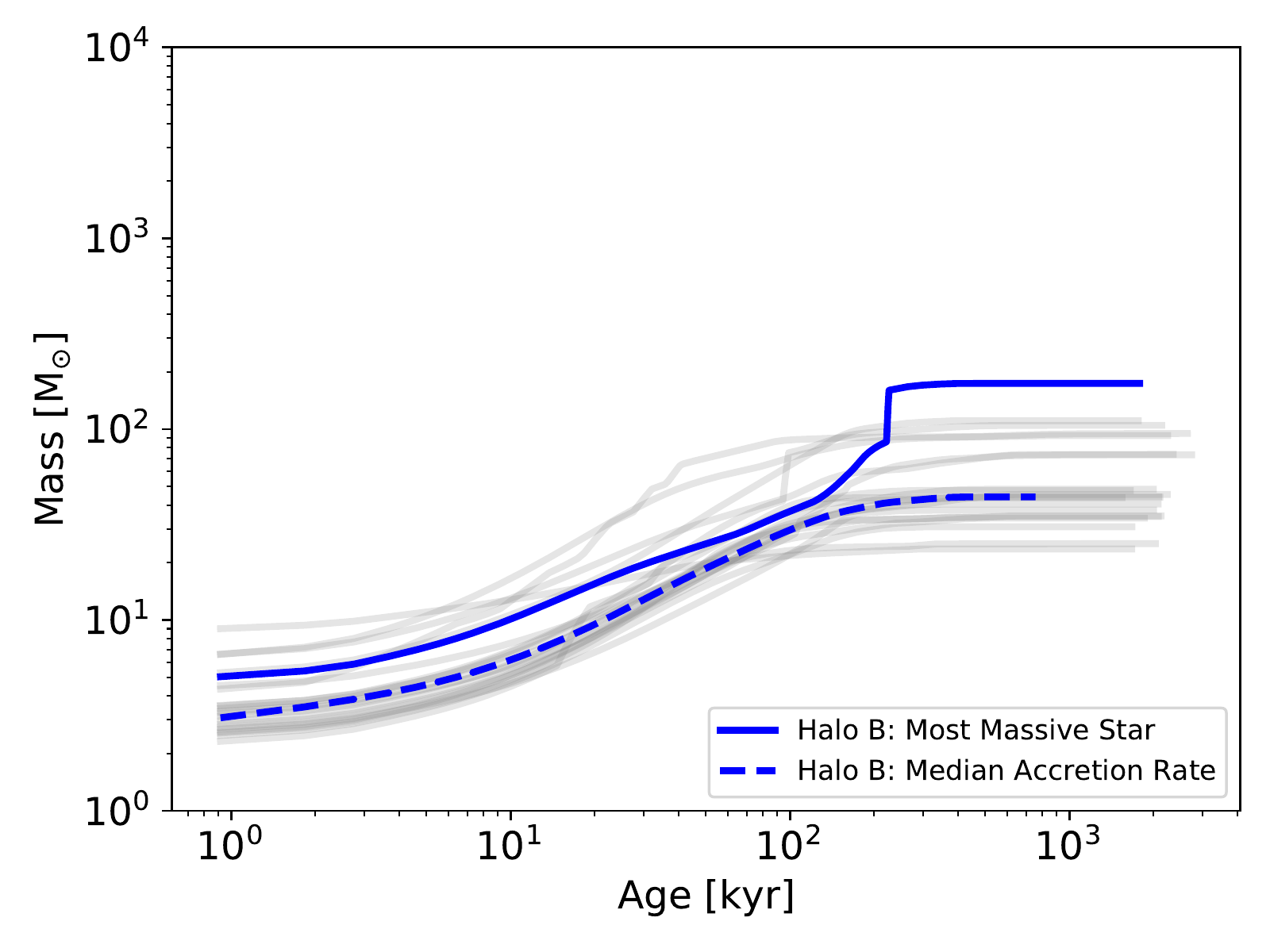}}
    \caption{
      The same as Figure \ref{Fig:HaloA_AccretionRates} except this time for the
      evolution of stars in \hbc. The most massive star in \hb has a mass of 173 \msolarc.
      Note that the sharp spike observed at approximately 200 kyr in the left hand panel
      is caused by a stellar collision.
    }
\label{Fig:HaloB_AccretionRates}
\end{center} \end{minipage}
\end{figure*}


\subsection{The onset and end of star formation}
Star formation is triggered when all of the criteria set out in \S \ref{Sec:StarFormation} are
fulfilled. In Figure \ref{Fig:ProjectionStart} we show a projection of the total gas number
density\footnote{Number density refers here to the total number density from each separate
  species weighted by each species atomic mass unit.} in the
regions surrounding the formation of the first star in each halo. The projections are made from the
first snapshot following star formation. Each panel is 2 pc across. In the left hand panel we
show the projection from \hac, which shows the formation of a single star (coloured in orange
at the centre of the green coloured gas cloud). The legend on the top left gives the mass and
age of the star at this time.
In the right-hand panel we show the projection for \hbc. In this case two stars are formed by
the time of the first
output following star formation. The most massive star has a mass of 37 \msolar and is 28 kyr old,
the second star has a mass of 6 \msolar and is 4 kyr old. Stars are coloured in red if the
accretion rate exceeds the SMS critical rate of 0.04 \msolaryrc, otherwise 
stars are coloured in blue (denoting PopIII stars). The most massive star is
coloured in orange. \\
\indent In Figure \ref{Fig:ProjectionEnd} we show the extent of star formation by the end of each
simulation. In the left panel we show the results for \ha and
for \hb in the right hand panel. Both simulations were terminated approximately 2 Myr 
after the formation of the first star. In both cased we terminate the simulations
before the imminent impact of chemical and mechanical feedback from the first supernovae.
At the end of the \ha simulation there are 99 stars in the simulation with masses ranging from
approximately 40 \msolar up to a maximum mass of over 6000 \msolarc.  Star formation in \ha
is widespread throughout the inner 20 pc of the halo with a number of different gas clouds
giving rise to star formation.
Furthermore, as the simulation develops the interactions between clouds triggers star formation as
individual clouds merge and interact. \hbc, on the other hand, contains essentially only a single
site of star formation due to the significantly smaller halo mass. At the end of the simulation of
\hb there are 21 stars with masses ranging from approximately 20 \msolar up to approximately
170 \msolarc.\\
\indent In Figure \ref{Fig:MassFunction} we show the mass distribution of stars at the final
output time. Stars from \ha are binned in blue, stars from \hb are binned in red. The median mass
of stars formed in \hb is 44 \msolar while the median mass of stars formed in \ha is 683 \msolarc.
The most massive star in \ha at the final output time is 6127 \msolar with another star having a mass
of 4477 \msolarc. \ha has 22 stars with masses exceeding 1000 \msolarc. The lowest mass star in
\ha has a mass of 40 \msolarc. The mass distribution of stars in \hb is significantly smaller, running
from 22 \msolar up to 173 \msolarc. However, it should be noted that due to our finite resolution
($\Delta x \sim 1000$ au) we cannot accurately probe the lower end of the initial mass function
of either halo and we are likely missing some lower mass stars. 
\begin{figure}
   \centering 
\includegraphics[width=0.5\textwidth]{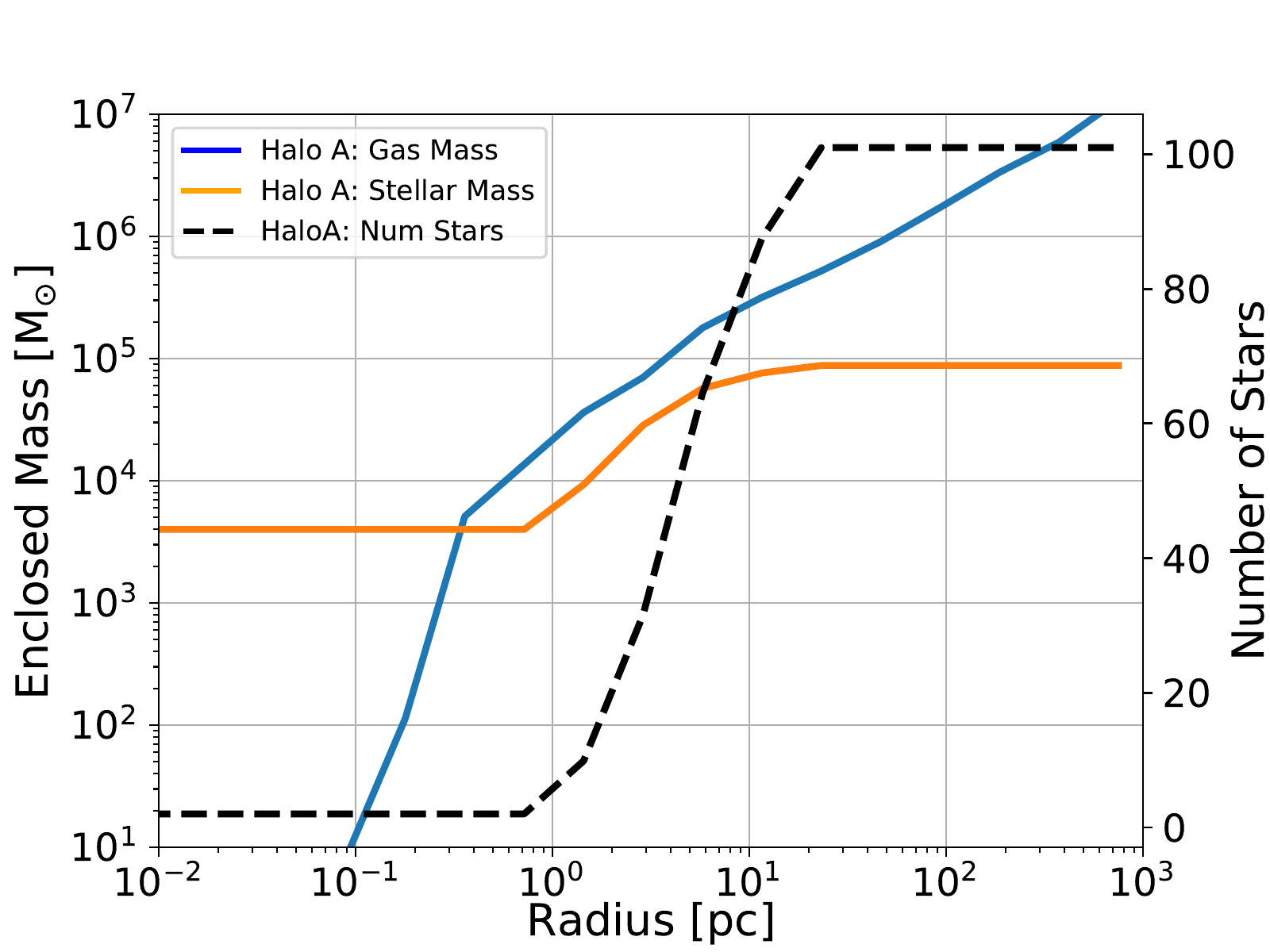}
\caption{The enclosed gas mass (blue line), enclosed stellar mass (orange line) and the number of
stars (dashed black line) as a function of radius. The radial profile is centred on the centre
of mass of stars in the simulation. The profile is created from the final dataset of \hac.
The stars are concentrated in the central 20 pc of the simulation with a total stellar mass of
approximately 90,000 \msolarc. The baryonic mass at 1 pc is approximately $3 \times 10^4$ \msolar
rising to approximately $6 \times 10^5$ \msolar at 20 pc. }
\label{Fig:EnclosedMass}
\end{figure}
\begin{figure*}
\centering
\begin{minipage}{175mm}      \begin{center}
\centerline{
    \includegraphics[width=18.0cm, height=12cm]{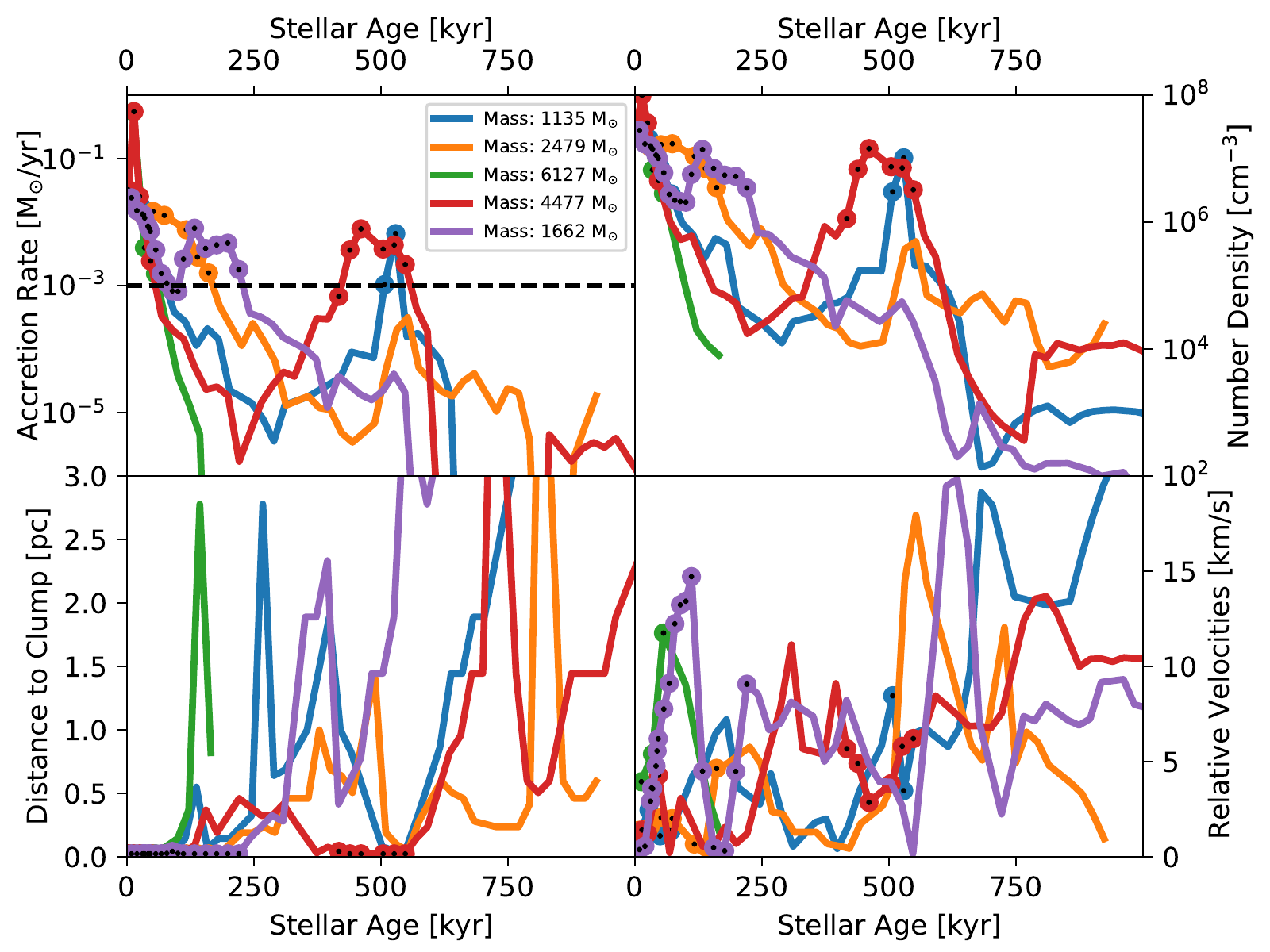}}
\caption{The evolution of the most massive star in \ha and four other stars which
  cross regions of high density during their evolution resulting in enhanced accretion rates.
  The accretion rate for each star (most massive in green) is shown in the
  top left panel. The final mass of each star ranges from approximately 1135 \msolar up to 6127
  \msolarc. Each of the stars (with the exception of the most massive star) shows a second
  spike of $\dot{M} \gtrsim 10^-3$ \msolaryr during its evolution. These times of high accretion
  are highlighted using circular markers with embedded black dots. For most stars this
  happens a few hundred kyr after formation. This spike in accretion is caused by the star
  crossing a region of high density.
  The gas number density surrounding each star is shown in the top right panel. The
  relative velocity of the stars (which also impacts accretion) is shown in the bottom
  right panel. Note that the relative velocity values can be as high as 10 - 15 \kms
  during a high accretion event as the star can be crossing through a dense clump.
  Finally, in the bottom left panel we plot the distance from each star to the
  nearest high density clump. Times of high accretion always corresponds to a distance
  of 0 pc from a clump. However, as stars cross the inner halo their distances from a high density
  clump varies in general. Note that for the most massive star, shown in green, the stellar
  age is quite short as this star formed near the end of the simulation.}
\label{Fig:Clumps}
\end{center} \end{minipage}
\end{figure*}
\subsection{The case for super-massive star formation}
In Figure \ref{Fig:ProjectionEnd} we saw that the mass of the most massive star in \ha was
6127 \msolarc. While this mass is well beyond the mass of ordinary PopIII stars
\citep{Turk_2009, Greif_2011, Wise_2012b, Crosby_2013, Susa_2014, Hirano_2014, Stacy_2016} and also more than 30 times
larger than the most massive star in \hb it is still well short of the mass often associated with
truly super-massive stars \citep[e.g.][]{Woods_2018} which are expected to have
end stage masses of $\sim 10^5$ \msolarc. Furthermore, \ha was chosen here to present near ideal
initial conditions in which to form a supermassive star.
To illustrate this point we show radial profiles of the halo properties of both \ha and \hb in
Figure \ref{Fig:RadialProfiles}. Figure \ref{Fig:RadialProfiles} shows the state of the gas
in each halo immediately prior to star formation. \ha is denoted by the blue line, \hb by the red line.
The temperature, \molH fraction, enclosed gas mass and mass inflow rate are shown in
clockwise order starting from the bottom left panel. \\
\indent \ha shows near isothermal collapse, although the gas does
show some degree of cooling at approximately 10 pc. The \molH fraction is very different between
the two haloes, with \ha showing a steep decline in \molH towards the centre due to the combination
of the LW background and the rapid assembly of mass. \hb on the other-hand shows a more typical
\molH evolution consistent with mini-halo formation. The enclosed mass plot in the top right panel
illustrates the different mass associated with each halo as a function of radius. \ha being well
inside the atomic cooling mass range. In the bottom right hand panel we see the mass inflow rate for
each halo. \ha shows mass infall rates averaging greater than 0.1 \msolaryr all the way into the
centre of the halo. \hbc's mass infall rates on the other hand fall from approximately
0.1 \msolaryr at 1 pc down to approximately $10^{-3}$ \msolaryr in the very centre. As we will see
this result is reflected in the initial accretion rates observed onto the protostars. \\
\indent In order to understand why, given apparent high mass infall rates that are greater than the
critical rate, a SMS does not form we need to examine the accretion rate that is measured onto the
stars within each simulation itself. In Figure \ref{Fig:HaloA_AccretionRates} we show the mass evolution and
the accretion rate onto every star in \hac, the most massive star in \ha is coloured in blue.
In the left panel we show the mass accretion rate while in the right hand panel we show the mass evolution.
In each case we plot as a function of the stellar age. The dashed line in the panel shows the mass evolution
of a star with the median\footnote{We define a
        median star here as the star at the end of the simulation which has a final mass
        closest to the median stellar mass at the end} stellar mass in the simulation.
The most massive star in \ha is formed
when two gas clouds within the central region merge and trigger star formation. The star quickly
grows in mass up to its final mass of 6127 \msolar within approximately 20 kyr, after which
its accretion rate declines. The shaded grey lines are the evolutionary tracks for each of the
other 100 stars in \hac. The trend is similar for every other star.\\
\indent Initially each star follows approximately the mass accretion rate as found in the radial profiles (see Figure \ref{Fig:RadialProfiles}).
However, in every case after a few tens of kyr the accretion rate onto each star drops dramatically and mass growth is halted.
This drop in accretion is not due to feedback since no ionising feedback
is present in these simulations. Instead mass accretion is terminated as the star accretes all the gas in
its surroundings and moves into a less dense region of the inner halo. All of the stars formed have a small initial velocity of a few \kms,
which it takes from the gas that initially formed the star. While this 
initial velocity is much smaller than the circular
velocity of the host halo it does cause the stars to move around and decouple from the high density gas cloud in which they initially formed. \\
\indent In Figure \ref{Fig:HaloB_AccretionRates} we show the mass evolution and accretion rate onto each star in \hbc.
  Again in the left panel we show the mass accretion rate while in the right hand panel we show the mass evolution.
  The mass evolution of the most massive star (coloured in blue) in \hb is a little different. In this case accretion is somewhat lessened
  at approximately $10^{-3}$ \msolaryr (left panel). The most
  massive star in \hb results from a stellar merger which then drives the star to increase its mass suddenly. This occurs
  when the main progenitor star is approximately 200 kyr old and results in a star of 173 \msolarc. However, the general
trends are similar with initial accretion rates always declining as stars move into less dense environments.\\

\subsection{Stellar Multiplicity}
\noindent In Figure \ref{Fig:EnclosedMass} we plot the enclosed gas (blue line) and
  stellar mass (orange line) of \ha as a function of radius.
The radial profiles are centred on the centre of mass of the stars in the halo. We also show the
number of stars (black dashed line) in each radial bin up to the number of stars in the halo (101). All of the stars
are concentrated in a radius of 20 pc of the halo centre. By the end of the simulation the total stellar mass
in \ha is 90,000 \msolar (cf. 1300
\msolar in \hbc). The star formation efficiency in the inner 20 pc (5 pc) of \ha (\hbc) is $\sim$ 23\%
($\sim 2$\%). The stellar to dark matter ratios are higher in each case since this 
inner regime is baryon dominated. 20 pc is the approximate Jeans length of the gas in \ha ($\sim 5$ pc for \hbc).
In order for the formation of a truly supermassive star in \ha all of the
stellar mass
(i.e. $\sim$90,000 \msolarc) would have needed to have been accreted onto the stellar surface. Instead
what happened was that the outer gas cloud in \ha fragmented into a small number of sub-clouds which
each generated hotspots of star formation. These sub-clouds tidally disrupt each other during the
subsequent evolution both triggering further star formation but also tidally disrupting accretion.
From a total baryonic reservoir of approximately 600,000 \msolar inside 20 pc, approximately 1\% of the baryons went
into the most massive star. To form a SMS
with a mass of close to 100,000 \msolar we would need over a fifth of the mass to 
flow into a single object. While this is possible it is clearly going to be 
very challenging given the turbulent nature of the environment in which these stars are forming. \\
\indent The stars in the central 20 pc region of \ha wander the inner parts of the halo in a random walk fashion. As
stars cross regions (filaments and clumps) of high density they can experience enhanced accretion rates. In Figure
\ref{Fig:Clumps} we show the environmental changes which occur for five stars within \hac. Four of the stars were
chosen because at some point in their evolution they undergo a second period of high accretion
($\dot{\rm{M}} > 1 \times 10^{-3}$ \msolaryrc) at least 100 kyr after their formation. We also include the
most massive star, coloured in green. Periods of high accretion 
are identified by a circular marker with an embedded black dot. 
Note that each of the stars (excluding the most massive) undergoes a ``bump'' in its accretion rate independent of the initial high accretion
rate value (top left panel of Figure \ref{Fig:Clumps}). The high accretion rate is directly correlated with the star
passing through a region of high density which is shown in the top right panel. The relative velocities of each of the stars
are shown in the bottom right panel while the distance from each star to the nearest high density clump is
shown in the bottom left panel. Average relative velocities are below 10 \kms{} with some spikes showing
velocities between 10 and 20 \kms. A high density clump is defined as any cell that has a density more than
10\% of the  Jeans Density. The distance of a star from a high density clump is highly variable with stars
encountering a high density clump only occasionally. Encounters
with filaments and clumps do occur however with an associated spike in
the accretion rate (and density) around the star.
What is also clear is that periods of high accretion are fleeting, lasting only a few kyr at most. Apart from these
short periods of rapid growth the initial mass of the
stars does not increase appreciably and a consistently high accretion rate is not maintained. This is similar in
spirit to what \cite{Smith_2018} found for accreting stellar mass black holes in similar high-z environments.\\
\indent In summary none of the stars in our simulation are able to sustain critical accretion rates
  for SMS formation. Nonetheless, the most massive star(s) in the halo will form of population of
  intermediate mass black holes (IMBHs) with masses ranging from 300 \msolar to several thousand \msolarc.
The subsequent
growth of these IMBHs within the halo then through 
mergers and accretion will then slowly build the black hole mass. 
The key issue preventing the formation of a monolithic SMS appears to be that
these haloes are highly turbulent and that sustaining accretion onto a single
object in such an environment is hugely challenging (see \cite{Chon_2017b} and \S
\ref{Sec:Discussion} for a more detailed discussion).
This conclusion is unlikely to be due to insufficient resolution: it is
well-known that low resolution damps turbulent motion very significantly
\citep[e.g.][]{Federrath_2010a, Downes_2012}. Thus insufficient resolution
will lead to an {\it over-estimate} of the likelihood of forming an SMS.

\begin{figure}
   \centering 
\includegraphics[width=0.5\textwidth]{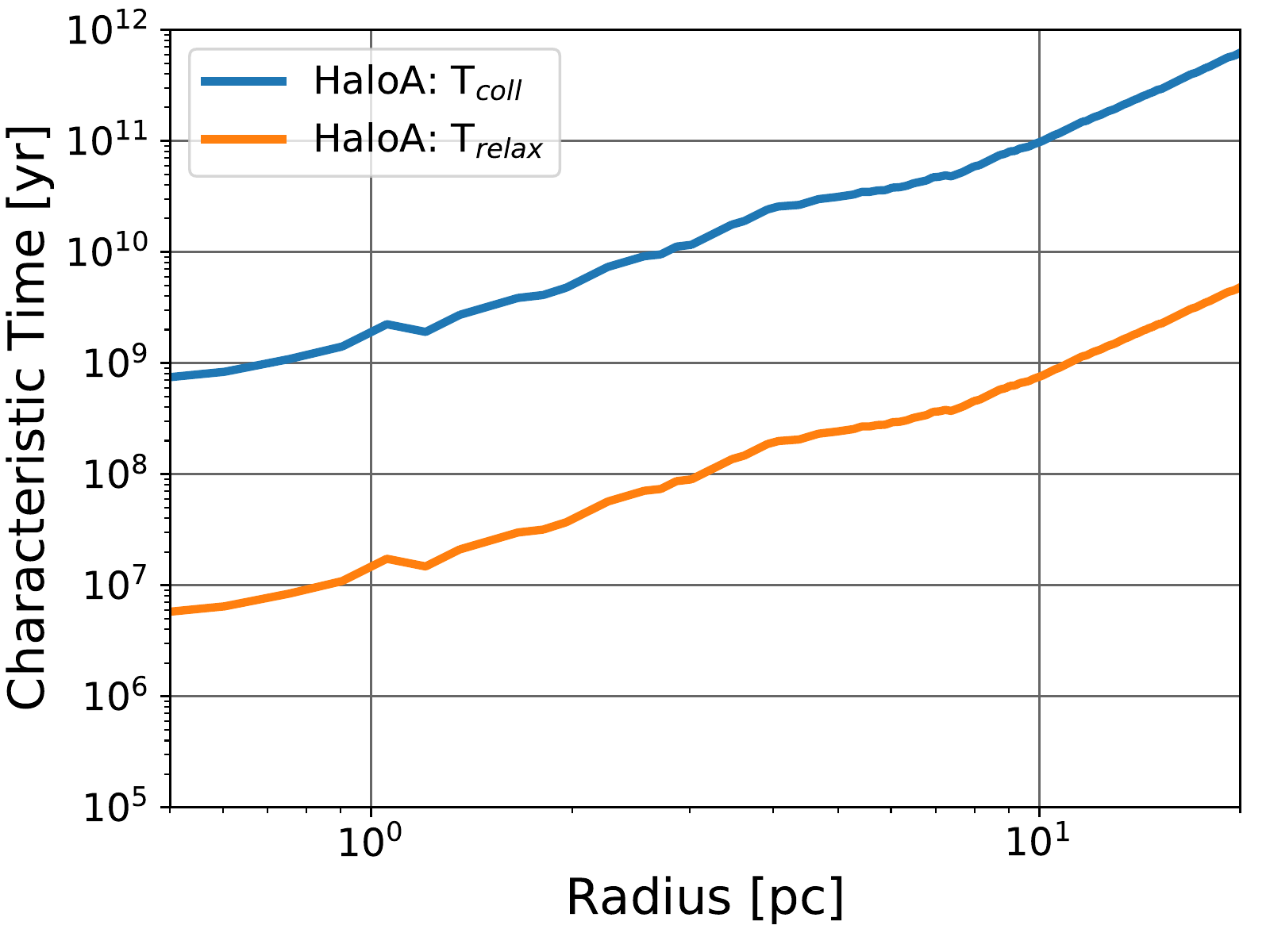}
\caption{The collisional time and relaxation times for the stars in \hac. The collisional times is
  an estimate of the time taken for a collision between two stellar populations. For the stars
  in \ha the time taken for a single collision to occur exceeds the Hubble time at all radii.
  The relaxation time is a measure of the time taken for the stellar system to contract to
  higher density, perhaps driving a higher collision rate, due to
  two body relaxation. For \ha the relaxation time is approximately 10 Myr at 1 pc. }
\label{Fig:Collisions}
\end{figure}

\section{Intermediate Mass Black Hole Merging and Implications for Gravitational Wave detection} \label{Sec:GW}
We begin this section by analysing the stellar content of \ha and in particular the likelihood
that \ha will undergo two-body collisional processes akin to the runaway collapse of a dense stellar
cluster \citep{PortegiesZwart_2004, Gurkan_2004, Freitag_2006, Regan_2009b, Katz_2015}. In Figure
\ref{Fig:Collisions} we plot both the collisional time for the stars in \ha and the relaxation
time of the stellar system in \ha as a
function radius. The collisional time, t$_{\rm{coll}}$, for a stellar population consisting of two
populations can be written as \citep{Freitag_2008}:
\begin{equation}
  t_{\rm{coll}} = 5 \ \rm{Gyr} \Big( {10^6 \over n_*} \Big ) \Big( {\sigma_{vel} \over 10 \ \rm{km \ s^{-1}}} \Big ) \Big ({2 R_{\odot} \over R_1 + R_2} \Big) \Big ({2 M_{\odot} \over M_1 + M_2} \Big)
\end{equation}
where $n_*$ is the stellar density, $\sigma_{vel}$ is the one dimensional velocity dispersion,
R$_1$ is the stellar radius of the first population of stars and R$_2$ is the stellar radius of the
second population. We choose the median stellar mass, 683 \msolarc, for M$_1$ and we choose our
maximum stellar mass, 6127 \msolarc, for M$_2$. The stellar radii are calculated using the standard
formulae from \cite{Stahler_1986}. The collisional time, plotted in blue, is a measure of time
required for one
stellar collision between these two populations. In orange we plot the relaxation time of the cluster.
The relaxation time is the time required for the system to contract due to two body interactions in a
populous cluster (N $\gg$ 10). The relaxation time is given by:
\begin{equation}
  t_{\rm{relax}} = 2 \ \rm{Myr} \Big ( {10 \over \lambda} \Big )  \Big( {10^6 \over n_*} \Big ) \Big( {\sigma_{vel} \over 10 \ \rm{km \ s^{-1}}} \Big )^3 \Big ( {1  M_{\odot} \over \langle m \rangle } \Big )^2
\end{equation}
where $\lambda = \ln (0.02 N)$ and  N is the number of cluster stars and $\langle m \rangle$
is the median stellar mass. The collisional time, t$_{\rm{coll}}$, is greater than the Hubble time
at all radii indicating that collisions are very unlikely for the stars in \hac. The relaxation time,
t$_{\rm{relax}}$, is approximately 10 Myr at 1 pc. However, this time exceeds the lifetime of the
stars in our cluster and contraction to very high densities appears unlikely and hence the
runaway collapse of the cluster is not predicted (or observed here). \\
\indent
At the end of the simulation \ha has 22 stars with masses greater than 1000 \msolar with
the most massive star having a mass of 6127 \msolarc. Within the central 1 pc of the 
  halo centre there is a total (stellar + gas) mass of $\sim 3 \times 10^4$ \msolarc.
  This puts \ha at the lower end of what defines a Nuclear Star Cluster (NSC) \citep[e.g.][]{Georgiev_2016}.
  However, subsequent, further rapid growth of this halo beyond this initial phase of star formation
  will drive the enclosed mass further upwards. This would have the effect of pushing the environment
  to a point where the central region of a galaxy becomes a proto-NSC. At this stage if the density
  of the core reaches a point where the enclosed total baryonic mass within the central parsec
  reaches approximately $10^6$ \msolar then core collapse can occur \citep{Davies_2011}
  potentially reaching a central black hole mass of M$_{BH} \sim 10^5$ \msolarc. Depending on the
  densities reached core collapse could occur on timescales of a few hundred Myr
  \citep{Davies_2011} although equally the NSC 
  could act as an incubator for the rapid growth of a seed black hole
  through gas accretion achieving
similar masses over a potentially longer timescale \citep{Natarajan_2020}. \\
\indent Our specific case here shows that we will have a  proto-galaxy that gets populated with a plethora of
weakly accreting black holes early in the proto-galactic evolution.
It is therefore reasonable to ask \textit{what are the prospects for the merger of this first generation
  of black holes into a population of more massive black holes?}. \\
\indent \cite{Pfister_2019} investigated the merging of black holes in high resolution
simulations ($\Delta x \sim 10$ pc) with black hole masses in the range
$10^4$ \msolar to $10^5$ \msolar in high-z galaxies. They found that dynamical friction from stars
has a more stabilizing effect on the black holes than the gas component but that the stellar
distribution in high-z galaxies is highly irregular and hence is not able to stabilise the black
hole orbits.
The irregular stellar distributions found by \cite{Pfister_2019} correlates with what
we see in our simulations here. \cite{Pfister_2019} found that black holes with masses less than
$10^5$ \msolar do not sink to the centre of a galaxy, but instead exhibit random walk characteristics.
However, their simulation neglected the impact of any radiative feedback from the black hole 
(which could impede the sinking timescale) and
also lacked sufficient resolution to detect high density gas which could promote dynamical friction. \\
\indent \cite{Toyouchi_2020}, building on previous investigations of \cite{Park_2017} and \cite{Park_2019},
investigated a more idealised scenario where they modelled the
dynamics of a single $10^4$ \msolar mass black hole incorporating the effects of dynamical friction from gas drag
and radiative feedback. They found that if the gas density is low that the black hole
does not sink to the centre - similar to what \cite{Pfister_2019} found. However, in the case where
the black hole encounters dense gas ($n_{gas} \gtrsim 10^6 \ \rm{cm}^{-3}$) the ram pressure of
the head wind (due to radiative feedback)
causes the black hole to lose orbital energy and so the black hole can sink to the centre on a
timescale of t$_{dyn} \sim$ 0.01 Myr, consistent with the dynamical friction timescale
in the Bondi-Hoyle-Littleton case (i.e. no radiative feedback),
and much shorter than the dynamical timescale in galactic gas disks.\\
\indent Taking the two results together this
appears to imply that the specific environment that the black hole finds
itself in will play a central role in determining the timescale for the black hole
to sink towards the centre. If the black hole can interact with sufficiently dense gas it may
sink towards the centre (even for black holes with masses of $\sim 10^4$ \msolarc) while in
the case where the black hole does not encounter sufficiently dense gas then the black hole
may wander the central region of the galaxy for many dynamical times. The galactic centres
in which our massive PopIII
stars form (especially \hac) contain a web of dense gaseous filaments (with gas clearly approaching
that critical level of $n_{gas} \gtrsim 10^6  \ \rm{cm}^{-3}$ (see Figure \ref{Fig:ProjectionEnd}) in the centre of the halo)
which may help to supply
the necessary ram pressure and extract orbital energy. A further build up of dense gas would potentially
help to promote efficient frictional forces even allowing for radiative feedback.

\subsubsection*{What then does this mean for potential gravitational wave detection?}

\begin{figure}
   \centering 
\includegraphics[width=0.525\textwidth]{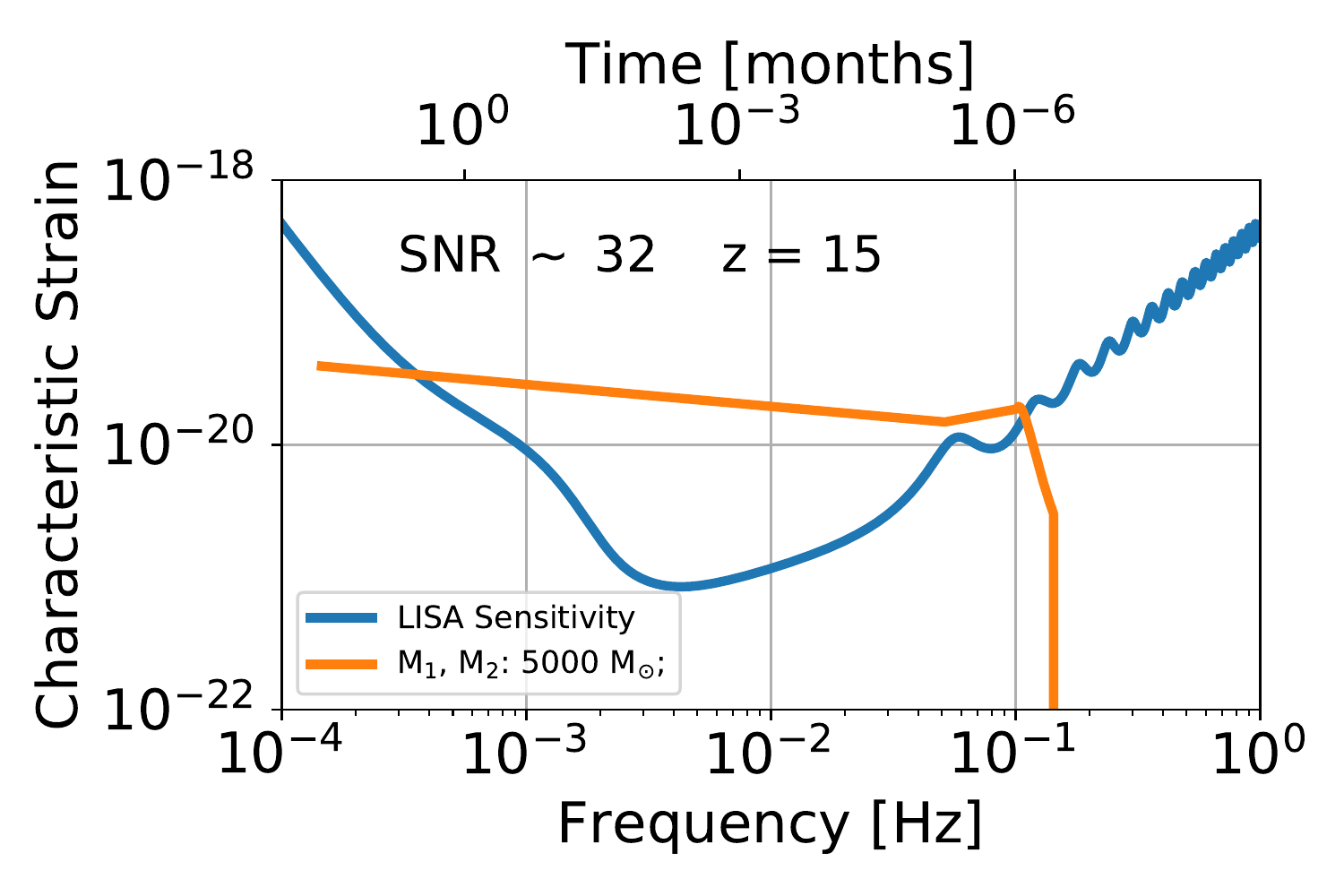}
\caption{The LISA sensitivity curve with the strain due to two massive black holes each with a mass
  of 5000 \msolarc. The merger of two 5000~\msolar black holes at $z = 15$ gives a
  signal to noise ratio of 32 that is visible by LISA. The inspiral will enter the LISA band at
  approximately $5 \times 10^{-4}$ Hz, corresponding to a time to merger of approximately 2 months.}
\label{Fig:Sensitivity}
\end{figure}

\noindent Within approximately 4 Myr after the formation of the first massive stars here there
will be a population of black hole seeds with masses in the range 300 \msolar (upper limit of pair
instability supernovae) up to 6000 \msolar
(and likely higher). In Figure \ref{Fig:Sensitivity} we plot the characteristic strain
expected from the merger of two black holes with masses of 5000 \msolar each at $z = 15$
we note that here that gravitational recoil effects can eject black holes from galactic
centres \citep[e.g.][]{Gultekin_2006}, however we consider the idealised case of a non-spinning
  binary black hole merger with a mass ratio, $q = 1$ and hence in this
  idealised case ejections
  can be avoided \citep{Holley-Bockelmann_2008, Morawski_2018, Dunn_2020}.
These masses are consistent with the masses of the stars found in this simulation and black holes
with similar masses are expected from the direct collapse of these massive PopIII
stars \citep{Heger_2003}. The merger of two black holes with these masses will produce a
gravitational wave signal detectable by LISA \citep{eLISA, Sesana_2016, Cornish_2020} emitting
gravitational waves between approximately $10^{-4}$ and $10^{-1}$ Hz (detector frame frequencies). The
Signal-to-Noise ratio (SNR) from the merger of two black holes with masses of 5000 \msolar at $z = 15$
reaches a SNR $\sim 32$ over its time inside the LISA band and hence is
well within LISA's detection parameters\footnote{The SNR was calculated
  using the LISA sensitivity calculator available at
  \url{github.com/eXtremeGravityInstitute/LISA_Sensitivity.git} \citep{Robson_2019}}.
The merger of a 10,000 \msolar binary black hole with a mass ratio of 1 at a redshift of 15
will enter the LISA band approximately 2 months before merger and complete
thousands of orbits before the final plunge. The SNR value is calculated by integrating
  over the entire frequency range over which the binary overlaps with the LISA band. The
SNR exceeds 10 approximately 12 hours before merger.
With this number of cycles LISA will be able to detect
the redshifted mass with a precision of close to 1\% \citep{Sesana_2013}, with the
  strongest signal and hence the highest SNR being achieved in the final few hours and
  hundreds of orbits \citep{Robson_2019}. The largest uncertainty
will however come from the determination of the redshift (or luminosity distance) which at this
redshift is expected to have an uncertainty of approximately 30\% \citep{Sesana_2013}. Additional
uncertainties due to weak lensing will also contribute at the percent level
\citep{Shapiro_2010, Petiteau_2011}
but the dominant error for these high-z sources will be LISA intrinsic errors on the measurement of
the wave amplitude. \\
\indent This redshift uncertainty will translate to a similar uncertainty in the chirp mass of the binary.
\cite{Klein_2016} modelled the merger of similar black hole masses and redshifts and using a
SNR = 7 finding that the expected number of detections would be 358 over a 5 year mission period
(their \texttt{N2A5M5L6} model in Table II). Their modelling only accounted for the merging of
heavy seed black holes following a galaxy merger and
neglected the in-situ models as described here which could increase the numbers of events potentially
discoverable by LISA significantly. This point was previously investigated by
\cite{Hartwig_2018} as a method to discriminate between seed formation scenarios. Further,
investigation of the number density of in-situ mergers and how they could influence the number
of detections by LISA is now underway (O'Brennan et al. in prep). The merger of less
  massive black holes ($\rm{M_{BH} \sim 1000}$ \msolarc) that are produced in these environments
  may also be detectable by 3rd generation, ground based, GW observatories. An exploration of their
  potential is outside the scope of this discussion but see \cite{Valiante_2020} for more details.
\section{Discussion and Conclusions} \label{Sec:Discussion}
The aim of this study was to examine the prospect of (super-)massive star formation in haloes
that are exposed to moderate LW radiation and also experience significant dynamical heating
effects due to mergers. We identified 79 such haloes in the Renaissance 
datasets \citep{Regan_2020} and then subsequently selected two haloes, which showed no star formation or metal enrichment prior to reaching the
atomic cooling limit, for re-simulation at enhanced resolution. These halo characteristics make these environments ideal for forming massive
stars. \\
\indent The two haloes, \ha and \hbc, were then targeted for re-simulation by 
increasing the resolution and by adding additional physics capabilities to provide a more self-consistent calculation of (super-)massive star formation in these haloes. Firstly, the spatial resolution was increased by a factor of 256 ($\Delta x \sim 1000$ au) and the mass resolution by a factor of
169 (M$_{\rm DM} \sim 170$ \msolarc). Additionally, self-shielding of \molH from LW was included and the Jeans length was refined by at least 64 cells at all times.  To model star formation within the collapsing gas clouds a star formation prescription was employed, which additionally tracks accretion
onto the stars and allows for the distinction between compact PopIII stars and massively inflated
SMSs. \\
\indent \ha was simulated with a LW radiation field impinging on it that followed exactly what was
found in the original simulations and so accounted for star formation from nearby galaxies. The
LW radiation field imposed is shown in Figure \ref{Fig:LWHistory}. \hb on the other hand was used as the
control halo and no LW field was imposed to test the impact that turning off the LW field would have.\\
\indent In the high-resolution re-simulation \ha formed 22 stars with masses greater than 1000
\msolarc. The most massive star in \ha has a mass of 6127 \msolarc. \hbc, re-simulated without
a LW radiation field, collapsed earlier in its evolution before reaching the atomic cooling limit.
The lack of any LW field allows PopIII star formation to occur prior to the onset of atomic line
cooling. In \hb a total of 21 stars
formed by the end of the simulation with the most massive star in \hb having a mass of 173 \msolarc.
While our results cannot be used to infer a PopIII initial mass function our results
are consistent with the results of other groups who have found that PopIII mini-haloes
tend to host a small number of stars with peak masses in the range of a few tens of
solar masses \citep{Hirano_2014, Susa_2014, Stacy_2016, Skinner_2020}.\\
\indent While the most massive star in \ha has a mass of more than 6000 \msolar its accretion rate had fallen
to less than $10^{-6}$ \msolaryr by the time the simulation terminated. No ionising radiative
feedback was employed in these simulations and hence radiative feedback is not the reason for
the falloff in accretion. Instead, while high accretion rates of greater than 0.1 \msolaryr initially
fall onto the protostar those rates are not sustained. The turbulent nature of the collapsing parent
cloud causes a multitude of sub-clouds to form and dissipate. The sub-clouds with masses of
thousands of solar masses form stars but also tidally disrupt other sub-clouds disrupting
accretion onto other (proto-)stars. This phenomenon is observed in all of the stars formed in the simulation.
On average stars accrete above 0.01 \msolaryr for less than 100 kyr before accretion is halted
due to external tidal disruption. \\
\indent What does this mean for the formation of SMSs and massive black hole seeds? This is only a
single halo simulated for 2 Myr. However, it is likely that the phenomenon of a
turbulent environment is common in the early stages of galaxy formation and
PopIII studies have also converged on the formation of multiple stars in mini-haloes
\citep[e.g.][]{Turk_2012}. It is therefore not surprising that this picture appears in
more massive galaxies as well. \cite{Chon_2016} and \cite{Chon_2017b} investigated a similar system
where, like here, they selected massive haloes from a larger scale parent simulation. Using
particle splitting techniques they preformed two very high resolution zoom-in simulations for
approximately 100 kyr each. In one case they found that the central disk fragmented into approximately
15 stars with masses of a few$\times 10^3$ \msolar - similar to what we find here
and indeed what was found in the more idealised simulation of \cite{Regan_2018b} where an external LW field was used to suppress PopIII star formation. In the
other case from \cite{Chon_2017b} they found much less fragmentation leading to an almost spherical collapse and the
formation of a handful of stars with masses in the range of a few $\times 10^4$ \msolarc.
They concluded that the tidal field surrounding the collapsing haloes in both cases played a major
role. Given that \ha in this study experiences a sharp rise in its impinging LW field from a
nearby star-bursting galaxies shortly before its own collapse similar tidal effects are also expected
here. We plan to evaluate the impact of tidal fields on the formation of haloes hosting massive
star formation in an upcoming study (Regan et al. in prep.). \\
\indent While the tidal effects can change the dynamics of the large scale disk, below the parsec
scale the tidal effects only act as a seed to further turbulence which grows through isothermal
collapse \citep{Chon_2017b}.
Turbulence is a common feature of present day giant molecular clouds in
which star formation is clearly evident \citep[e.g.][]{Girichidis_2020, Lee_2020, Krause_2020}
and while in both cases the gas is moving super-sonically the coolants available to the gas in both
cases make direct comparisons difficult. 
After approximately 2 Myr the stars in \ha will start to directly
collapse in black holes retaining close to 100\% of the stellar mass \citep{Heger_2003}. While this
event is unlikely to have any immediate impact on the mass accretion rate onto the black hole it will
leave this halo with a small population of massive black holes only a few Myr after the onset of galaxy formation. Allowing the galaxy to evolve further  (through a few tens of dynamical times) may
reveal than these intermediate mass black holes can sink towards the centre of the halo and subsequently merge to form more massive black holes with masses close to $10^5$ \msolar (similar to what most canonical galaxy formation
simulation use as their seed mass). Such black hole mergers may even be
detectable by LISA as we demonstrate in Figure \ref{Fig:Sensitivity}. However, it is likely that some of this population of IMBHs will not merge or accrete any significant amount of gas, will display very low duty cycles and will wander their parent galaxy \citep{Tremmel_2018, Reines_2020, Barausse_2020}.\\
\indent To develop this model further what will actually be required is further high
resolution simulations which can model the formation of massive stars in high-infall rate haloes,
perhaps independent of metallicity \citep{Chon_2020, Regan_2020a}, which experience
a large transfer of baryonic mass towards their centres. These simulations will equally need to
be able to differentiate between PopIII star formation and SMS star formation, as done here, but
for a longer time period and beyond the formation of the subsequent black holes.
Tracking the evolution of these truly seed black holes is the next frontier. 

\section*{Acknowledgments}

\noindent JR acknowledges support from the Royal Society and Science Foundation Ireland under
grant number URF$\backslash$R1$\backslash$191132.
JHW is supported by National Science Foundation grants AST-1614333 and
OAC-1835213, and NASA grants NNX17AG23G and 80NSSC20K0520.  
BWO acknowledges support from NSF grants PHY-1430152, AST-1514700, AST-1517908, and  OAC-1835213, by NASA grants NNX12AC98G and NNX15AP39G, and by HST-AR-13261 and HST-AR-14315.
JR wishes to acknowledge the DJEI/DES/SFI/HEA Irish Centre for High-End Computing (ICHEC) for the
provision of computational facilities and support on which the zoom simulations were run.
The authors also thank the Extreme Gravity Institute at Montana State University for making the
LISA sensitivity curve scripts available on github. 
The original Renaissance simulations were performed on Blue 
Waters, which is operated by the National Center for Supercomputing Applications (NCSA)
with PRAC allocation support by the NSF (awards ACI-1238993, ACI-1514580, and OAC-1810584).
This research is part of the Blue Waters sustained-petascale computing project, which
is supported by the NSF (awards OCI-0725070, ACI-1238993) and the state of
Illinois. Blue Waters is a joint effort of the University of Illinois at
Urbana-Champaign and its NCSA.  The freely available plotting library {\sc
matplotlib} \citep{matplotlib} was used to construct numerous plots within this
paper. Computations and analysis described in this work were performed using the
publicly-available \enzo{}\citep{Enzo_2014, Enzo_2019} and \yt{} \citep{YT} codes,
which are the product of a collaborative effort of many independent scientists
from numerous institutions around the world. Their commitment to open science
has helped make this work possible. Finally, the authors thank the anonymous referee
for a detailed and very helpful report.

\label{lastpage}
\bibliographystyle{mn2e}
\bibliography{mybib}
\end{document}